\documentclass[11pt]{article}
\pdfoutput=1

\usepackage[usenames,dvipsnames,svgnames,table]{xcolor} 
\usepackage[obeyspaces,hyphens,spaces]{url}
\usepackage{jcapmod}

\usepackage{shorthand}
\usepackage{mathtools}
\usepackage{booktabs}
\usepackage[english]{babel}
\usepackage{amsmath,amssymb,amsbsy,amstext, amsthm, simplewick, amsfonts,braket}
\usepackage{hyperref}
\usepackage{graphicx}
\usepackage[small]{caption}
\usepackage{slashed}
\usepackage[separate-uncertainty=true,multi-part-units=single]{siunitx}
\usepackage{upgreek}
\usepackage{framed}
\usepackage{wrapfig}
\usepackage{multirow}
\usepackage{bbm}

\usepackage[utf8x]{inputenc}
\usepackage{selinput}
\usepackage{bm}
\usepackage{float}
\usepackage{geometry}
\usepackage{yfonts}
\usepackage{caption}
\usepackage{subcaption}
\usepackage{sidecap}
\usepackage{longtable}
\usepackage{anyfontsize}
\usepackage{dsfont}
\usepackage{tikz}
\usepackage{relsize}
\usepackage{tcolorbox}

\usepackage{colortbl}
\definecolor{lightgreen}{cmyk}{0.2, 0, 0.2, 0.2}
\definecolor{lightgray2}{cmyk}{0.1,0.1,0,0.1}
\definecolor{Red2}{RGB}{214, 39, 40}
\definecolor{Blue2}{RGB} {31, 119, 180}
\definecolor{Orange2}{RGB}{255, 127, 14}
\definecolor{Green2}{RGB}{44, 160, 44}
\definecolor{greyish2}{rgb}{.96,.96,.96}

\makeatletter
\newlength{\apb@width}
\newcommand{\autoparbox}[2][c]{\settowidth{\apb@width}{#2}\parbox[#1]{\apb@width}{#2}}

\makeatother


\allowdisplaybreaks[1]
\definecolor{darkgreen}{RGB}{50,150,0}

\definecolor{bluegreen}{RGB}{10,150,150}

\newcommand{\ex}[1]{\langle #1 \rangle}
\renewcommand{\v}[1]{ \mathbf{#1} }
\newcommand{\bfk}{{\bf  k}}

\newcommand{\bfx}{{\bf  x}}

\def\disc {\text{Disc}}

\def\beq{\begin{equation}}
\def\eeq{\end{equation}}

\begin{document}


\newgeometry{top=2cm, bottom=2cm, left=2cm, right=2cm}

\pagenumbering{roman}
\begin{titlepage}
\baselineskip=15.5pt \thispagestyle{empty}

\begin{center}
{\fontsize{19}{24}\selectfont  \bfseries  Snowmass White Paper:}\\[8pt] 
{\fontsize{22}{24}\selectfont  \bfseries The Cosmological Bootstrap}
\end{center}

\vspace{-0.2cm}
\begin{center}
{\fontsize{12}{18}\selectfont Daniel Baumann,$^{1,2,3}$ Daniel Green,$^4$ Austin Joyce,$^5$ Enrico Pajer,$^{6}$\\[4pt] Guilherme L. Pimentel,$^{7,1}$ Charlotte Sleight,$^{8}$ and Massimo Taronna$^{9,10,11}$} 
\end{center}

\begin{center}
     \vskip 3pt
\textit{$^1$ Institute of Physics, University of Amsterdam, Amsterdam, 1098 XH, The Netherlands}

  \vskip 3pt
\textit{$^2$  Center for Theoretical Physics, National Taiwan University, Taipei 10617, Taiwan}

  \vskip 3pt
\textit{ $^{3}$ Physics Division, National Center for Theoretical Sciences, Taipei 10617, Taiwan} \\

  \vskip 8pt
\textit{$^4$  Department of Physics, University of California at San Diego, La Jolla, CA 92093, USA}\\

  \vskip 8pt
\textit{$^5$  Kavli Institute for Cosmological Physics, Department of Astronomy and Astrophysics,\\
University of Chicago, Chicago, IL 60637,  USA}\\

  \vskip 8pt
\textit{$^6$  Department of Applied Mathematics and Theoretical Physics, \\
University of Cambridge,
Wilberforce Road, Cambridge, CB3 0WA, UK}\\

\vskip 8pt
\textit{$^7$ Lorentz Institute for Theoretical Physics, Leiden University, Leiden, 2333 CA, The Netherlands}\\

\vskip 8pt
\textit{$^8$ Centre for Particle Theory and Department of Mathematical Sciences,\\
Durham University, Durham, DH1 3LE, UK}\\

\vskip 8pt
\textit{$^9$ Dipartimento di Fisica “Ettore Pancini”, Universit\`a degli Studi di Napoli Federico II,\\
Monte S. Angelo, Via Cintia, 80126 Napoli, Italy}\\

\vskip 3pt
\textit{$^{10}$ Scuola Superiore Meridionale, Universit\`a degli Studi di Napoli Federico II,\\
Largo San Marcellino 10, 80138 Napoli, Italy}\\

\vskip 3pt
\textit{$^{11}$ INFN, Sezione di Napoli, Monte S. Angelo, Via Cintia, 80126 Napoli, Italy}
\end{center}

\begin{center}{\bf Abstract}
\end{center}
\noindent
This white paper summarizes recent progress in the cosmological bootstrap, an approach to the study of the statistics of primordial fluctuations from consistency with unitarity,
locality and symmetry assumptions. We review the key ideas of the bootstrap method,
with an eye towards future directions and ambitions of the program. Focusing on recent progress involving de Sitter and quasi-de Sitter backgrounds, we highlight the role of singularities and unitarity in constraining the form of the correlators.  We also discuss nonperturbative formulations of the bootstrap, connections to anti-de Sitter space, and potential implications for holography.


\vspace{1.2cm}
\hrule 
\begin{center}
Submitted to the  Proceedings of the US Community Study\\ 
on the Future of Particle Physics (Snowmass 2021)
\end{center}
\hrule

\end{titlepage}
\restoregeometry

\thispagestyle{empty}
\setcounter{page}{2}
\tableofcontents

\newpage
\pagenumbering{arabic}
\setcounter{page}{1}

\clearpage

\setcounter{page}{3}
\section{Introduction}

The statistics of the primordial density fluctuations offer a unique opportunity to probe the earliest moments of the universe~\cite{Hu:1996yt,Spergel:1997vq,Dodelson:2003ip}.  The seeds of all structure are believed to have been created in a phase of cosmic inflation~\cite{Guth:1980zm,Linde:1981mu,Albrecht:1982wi}, an era where both quantum mechanics and gravity play an essential role.\footnote{See the Snowmass white paper~\cite{Snowmass2021:inflation} for a summary of the current theoretical and observational status of inflation.}
During the inflationary period, small quantum fluctuations were stretched to cosmological distances, rippling the fabric of spacetime in an apparently random, but correlated fashion~\cite{Mukhanov:1981xt,Hawking:1982cz,Guth:1982ec,Starobinsky:1982ee,Bardeen:1983qw}. These correlations retain a memory of their genesis, providing us a rare glimpse of the universe in its infancy.

\vskip 4pt
An intriguing feature of inflation is that our view of this epoch is frozen in time. We can only make inferences about the inflationary era from spatial correlations in the initial conditions for the post-inflationary universe. 
These primordial correlations live on the future boundary of the inflationary spacetime or, equivalently, the past boundary of the hot Big Bang universe. The detailed structure of these boundary correlations encodes information both about the dynamics and particle content of inflation. Time does not appear explicitly in the observed correlations, but is instead encoded in their scale dependence, because modes of different wavelength freeze out at different times during inflation. By measuring this shape dependence of the late-time cosmological correlations, we hope to infer the physics of the inflationary era.

\vskip 4pt
The standard approach to make predictions for the inflationary correlations is to follow the evolution of these correlations through the entirety of the spacetime---from their origin as quantum fluctuations until they imprint themselves at reheating~\cite{Starobinsky:1985ibc,Salopek:1990jq,Sasaki:1995aw,Maldacena:2002vr,Weinberg:2005vy}. This approach has the desirable feature that it makes certain aspects of the physics---such as locality, causality and unitarity---completely manifest.  
However, this comes at a price: ensuring these properties at every moment in time requires us to perform difficult time integrals over all of the inflationary evolution. Despite these challenges, many heroic computations have been carried out~\cite{Maldacena:2002vr,Creminelli:2003iq,Creminelli:2004yq,Weinberg:2005vy,Seery:2005wm,Chen:2006nt,Chen:2006xjb,Chen:2006dfn,Seery:2006vu,Seery:2008ax,Leblond:2008gg,Chen:2009zp, Arroja:2009pd,Chen:2009bc,Adshead:2009cb}, and by now a rich spectrum of inflationary phenomenology is known~\cite{Arkani-Hamed:2003juy,Alishahiha:2004eh,Bartolo:2004if,Dimopoulos:2005ac,Creminelli:2006xe,Cheung:2007st,Cheung:2007sv,Holman:2007na,Weinberg:2008hq,Chen:2008wn,Chen:2009we,Green:2009ds,Baumann:2009ds,Chen:2010xka,Adshead:2011jq,Baumann:2011su,Endlich:2012pz,Namjoo:2012aa,Martin:2012pe, Baumann:2014nda,Wands:2007bd,Snowmass2021:CosmoEFT}.

\vskip 4pt
The {\it cosmological bootstrap} is a complementary strategy that is motivated by two important features of inflationary cosmology: First, as described above, we can't observe the time evolution during inflation directly, but instead have to {\it infer} it from the static boundary correlations. Second, inflation is likely to have occurred at energies far exceeding those probed by particle experiments, where our knowledge of the correct theory of particle physics is highly uncertain.
A conservative approach to both of these challenges is to 
 directly reconstruct ({\it bootstrap}) the cosmological correlations  on the late-time reheating surface, using cherished physical principles like {\it locality}, {\it unitarity}, and {\it symmetries} as consistency requirements~\cite{Maldacena:2011nz,Raju:2012zr,Raju:2012zs,Raju:2011mp,Mata:2012bx,Bzowski:2011ab,Bzowski:2012ih,Bzowski:2013sza,Bzowski:2019kwd,Kundu:2014gxa,Kundu:2015xta,Arkani-Hamed:2015bza,Shukla:2016bnu,Arkani-Hamed:2017fdk,Arkani-Hamed:2018kmz,Baumann:2019oyu,Sleight:2019mgd,Sleight:2019hfp,Green:2020whw,Pajer:2020wxk,Baumann:2020dch,Sleight:2020obc,Sleight:2021iix,Baumann:2021fxj,Meltzer:2021zin,Hogervorst:2021uvp,DiPietro:2021sjt,Sleight:2021plv}. 
 The bootstrap method then asks what space of correlations is consistent with these basic physical conditions. In many cases, these combined requirements are so constraining as to uniquely define the answer, reproducing the results of detailed bulk calculations and enabling even more complex ones.
 This strategy of focusing directly on observables and their consistency 
has yielded numerous insights into the structure of anti-de Sitter space (conformal field theories)~\cite{Poland:2016chs,Hartman:2022zik} and flat space (scattering amplitudes)~\cite{Elvang:2020lue,Kruczenski:2022lot}, which provided inspirational success stories.  Here, we will review recent progress in applying these ideas to cosmology.
 
\vskip 4pt
Ideally, one would like to classify all possible patterns of primordial fluctuations based on general principles.  However, as a practical starting point, it is useful to assume that the near scale invariance~\cite{Akrami:2018odb} and Gaussianity~\cite{Akrami:2019izv} of the observed fluctuations are tied to more fundamental principles.\footnote{See e.g.~\cite{Slosar:2019gvt,Meerburg:2019qqi} for discussions of models of inflation that violate these assumptions.}  This means that we take the observed scale invariance to imply
an approximate symmetry of the dynamics
and focus on low-order correlation functions beyond the Gaussian approximation. 
For simple processes involving three- and four-point functions of the light particles sourced during inflation---adiabatic density fluctuations and primordial gravitational waves---these correlation functions can be calculated using standard perturbative methods adapted to the cosmological setting. However, for even slightly more complicated physical processes, standard calculations are often intractable, motivating the search for new calculational approaches.

\vskip 4pt
The cosmological bootstrap program aims to further illuminate both conceptual and observational questions about the very early universe.   We only observe one universe and therefore we depend on theory to connect individual signatures to deeper characteristics of the inflationary era. The cosmological bootstrap provides a precise map between the spectrum of particles, their interactions, and their (quantum) state, to the analytic structure of the correlators.  These very properties of the correlators can also be crucial in the search of primordial non-Gaussianity observationally~\cite{Creminelli:2004yq,Dalal:2007cu,Baumann:2021ykm}.  In addition, the self-consistency of the correlators with unitarity and locality can also limit the range of parameter space beyond what is naively allowed within effective field theory~\cite{deRham:2022hpx} (e.g.~through positivity bounds on EFT coefficients~\cite{Pham:1985cr,Adams:2006sv}).  A natural hope of the bootstrap program is that it will continue to unearth new and unexpected connections between fundamental principles and correlators, theory and data.  

\vskip 4pt
Understanding the perturbative structure of correlators from first principles is also an important step toward answering a number of conceptual questions about inflation and/or de Sitter space.  Infrared divergences have long plagued loop calculations in cosmology and cast doubt on our understanding of the inflationary epoch~\cite{Ford:1984hs,Antoniadis:1985pj,Tsamis:1994ca, Tsamis:1996qm,Tsamis:1997za,Polyakov:2007mm,Polyakov:2009nq,Senatore:2009cf,Giddings:2010nc,Giddings:2010ui,Burgess:2010dd,Marolf:2010nz,Krotov:2010ma,Marolf:2010zp,Rajaraman:2010xd,Marolf:2011sh,Giddings:2011zd,Giddings:2011ze,Senatore:2012nq,Pimentel:2012tw,Senatore:2012ya,Polyakov:2012uc,Beneke:2012kn,Akhmedov:2013vka,Anninos:2014lwa, Akhmedov:2017ooy,Hu:2018nxy,Akhmedov:2019cfd}.  Recent progress has been made in understanding aspects and, in some cases, resolve these divergences~\cite{Gorbenko:2019rza,Baumgart:2019clc,Mirbabayi:2019qtx,Cohen:2020php,Mirbabayi:2020vyt,Baumgart:2020oby,Cohen:2021fzf}. These results must ultimately connect to the bootstrap, and offer both a test of these direct calculations, while inspiring new insights into the structure of these correlators.

\vskip 4pt
In this white paper, we review recent developments in the cosmological bootstrap in a number of directions.
In Section~\ref{sec:Symmetries}, we summarize the basic properties of cosmological correlators in perturbation theory, including their definition, singularity structure, and symmetries. We then review how these properties can be used to bootstrap inflationary observables both when interactions preserve and violate de Sitter boost symmetry.
In Section~\ref{sec:Unitarity}, we describe the constraints imposed by unitarity on cosmological correlations, both perturbatively, where it manifests itself in the form of cutting rules, and nonperturbatively as in the positivity of spectral densities.
In Section~\ref{sec:AdS}, we summarize how insights from the physics of anti-de Sitter space can be imported into the de Sitter context and comment on the obstacles to formulate de Sitter holography.
Finally, in Section~\ref{sec:Conclusions}, we list some important challenges and opportunities for the future.

\section{Symmetries and Singularities}
\label{sec:Symmetries}
To construct a cosmological bootstrap, we need to understand how the physical inputs of locality, unitarity, and symmetry constrain the observable outputs of inflation. We first review the definition of the boundary objects of interest---the cosmological wavefunction and boundary correlators. We then describe how symmetries and locality constrain these objects, and describe their structure of singularities.
(We discuss constraints from unitarity in Section~\ref{sec:Unitarity}.) We then summarize some examples of how one can use these inputs to bootstrap inflationary correlators both in situations where de Sitter symmetries control the dynamics, and in cases where interactions are sensitive to the departure from exact de Sitter space.

\subsection{Correlators in an Inflationary Universe}\label{sec:correlators}

Under relatively mild assumptions the fluctuations in the matter density of the late universe can be traced back in time to the beginning of the hot Big Bang. Moreover, if inflation is correct, then the initial surface of the hot Big Bang is identified with the final surface of an approximate de Sitter spacetime and correlations on this surface are the fundamental cosmological observables. As typically only the light fields survive until the end of inflation, our main interest is the correlations of these degrees of freedom (possibly sourced by the interactions with massive degrees of freedom in the bulk spacetime).\footnote{From a more abstract viewpoint, understanding the properties of general correlation functions in cosmological spacetimes is important, as it provides valuable insights into the structure of QFT in curved spacetimes, which still holds many deep mysteries.} Every inflationary model has two compulsory massless degrees of freedom: the Goldstone boson of broken time translations~\cite{Creminelli:2006xe,Cheung:2007st} and the graviton. The former sources density fluctuations in the late universe and is also called the adiabatic mode. 
In the following, we will define the inflationary boundary correlators more precisely and then show how they can be bootstrap using knowledge of their symmetries and singularities.

\subsubsection{Back to the Future}
\label{subsubsec::BttF}

\begin{figure}[t]
    \centering
    \includegraphics[scale=.8]{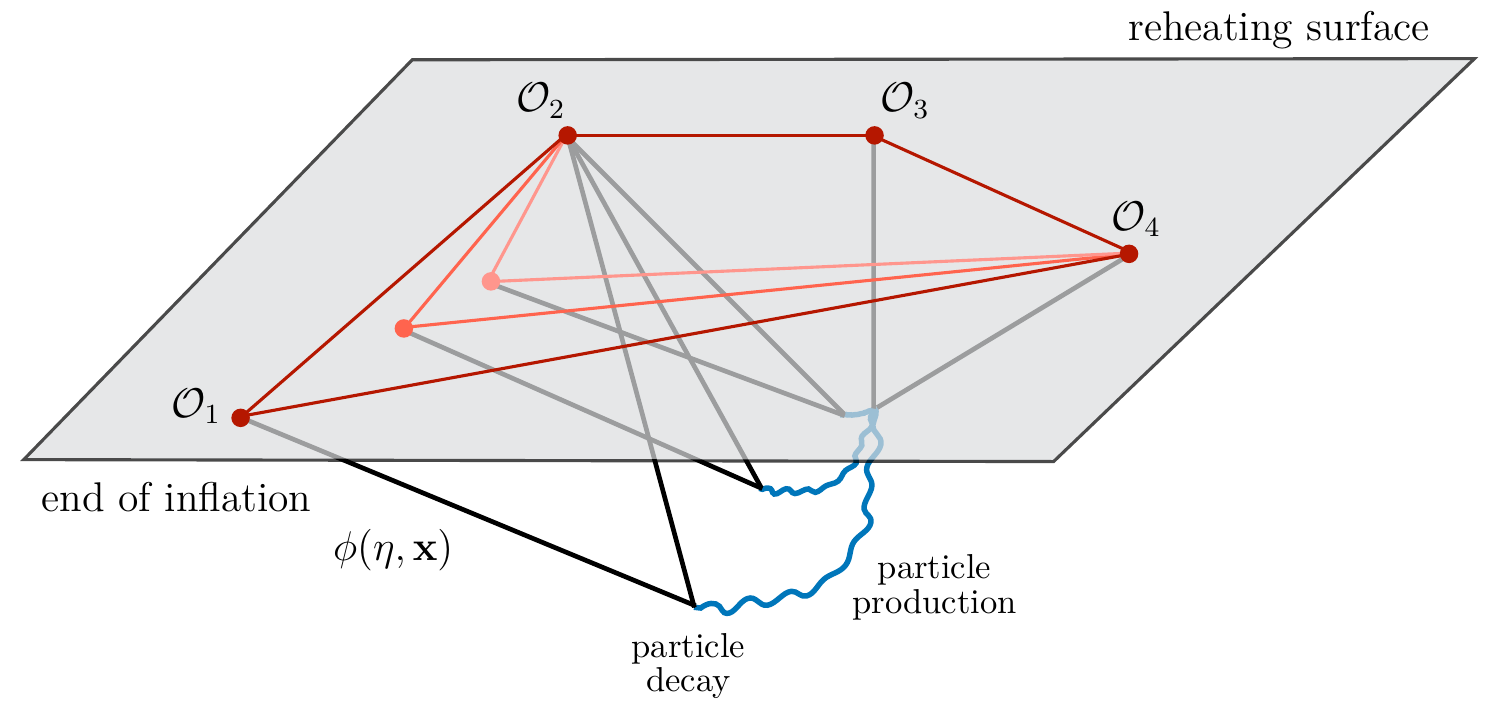}
    \caption{The pattern of correlations measured {\it after} inflation traces the dynamics and retains a memory of the physics {\it during} inflation. Because fluctuations of different wavelength freeze out at different times, the scale/shape dependence of late-time correlations encodes the time-dependant inflationary physics in a purely static object. An aspect of this remarkable connection is that we can learn about the particle content present during inflation by measuring subtle correlations imprinted in correlations on the reheating surface.
    }
    \label{fig:boundary}
\end{figure}

\noindent
The natural observables in cosmology are different from those in flat space (or even in anti-de Sitter space). In flat space, we are interested  
in the (squares of) transition amplitudes between asymptotic states:~the S-matrix. In the cosmological context, we don't have the luxury of specifying the states of interest in both the far past and future. Instead, we can only specify the initial conditions, which are then evolved forward in time. As a consequence, the observables in cosmology are correlation functions evaluated in this initial state~\cite{Maldacena:2002vr,Weinberg:2005vy}
\begin{equation}
\langle {\cal O}_1(\eta_1,\bfx_1) \cdots {\cal O}_n(\eta_n,\bfx_n) \rangle = \langle in |{\cal O}_1(\eta_1,\bfx_1) \cdots {\cal O}_n(\eta_n,\bfx_n) | in \rangle\,,
\end{equation}
which are called {\it in-in} correlators. It is the late-time value of these correlation functions, evaluated on the future boundary $\eta\to 0$, that is of interest for the subsequent cosmological evolution of the universe.
The homogeneity of spatial slices in cosmology makes it convenient to consider correlation functions in Fourier space, which we will do in the following. 

\vskip6pt
\noindent
{\bf The cosmological wavefunction}\\[2pt]
\noindent
The fact that we are not able to specify both $\lvert{\it in}\rangle$ and $\lvert{\it out}\rangle$ states in cosmology  motivates us to try to describe the $\lvert {\it in}\rangle$ state more explicitly, as it, in principle, contains the same information as all possible correlation functions evaluated in this state (though rearranged in a nontrivial way). The wavefunctional is not directly observable, but its somewhat more primitive nature makes it a simpler object than correlation functions.\footnote{In this respect, it is somewhat like the S-matrix, which itself is not directly observable, but from which physical observables are relatively easy to extract.}

\vskip 4pt
The state of interest is the (interacting) vacuum state processed by cosmological evolution. It can be represented in the basis of (Heisenberg picture) eigenstates of fields in the theory, $\lvert \varphi\rangle$, as $\Psi[\varphi,\eta] = \langle \varphi\rvert {\it in}\rangle$,
where the field eigenstates satisfy $\phi(\eta,\bfx)\lvert \varphi\rangle = \varphi(\bf x)\lvert \varphi\rangle$. The wavefunctional provides a probability distribution for spatial field configurations at time $\eta$.  

\vskip 4pt
In perturbation theory, it is convenient to parameterize the wavefunctional as
\begin{equation}
\log\Psi[\varphi, \eta] = \sum_{n = 2}\frac{1}{n!}\int \frac{{\rm d}^{d}k_1\cdot\cdot\cdot {\rm d}^{d}k_n}{(2\pi)^{3n}}\,\varphi_{\bfk_1}\cdot\cdot\cdot \varphi_{\bfk_n}\,(2\pi)^3  \delta(\bfk_1+\cdots+\bfk_n)\, \psi_n(\bfk_N)\,,
\label{eq:wfcoeff}
\end{equation}
where $\psi_n(\bfk_N)$ are the {\it wavefunction coefficients} expressed in Fourier space, with
$\bfk_N \equiv \{\bfk_1, \ldots, \bfk_n \}$. 
The late-time wavefunctional has a path integral representation
\begin{equation}
 \Psi[\varphi, \eta]  ~=
\int\limits_{\substack{\hspace{-0.3cm}\phi(\eta) \,=\,\varphi\\ \hspace{-0.7cm}\phi(-\infty)\,=\,0}} 
\hspace{-0.35cm} \raisebox{-.05cm}{ ${\cal D} \phi\, e^{iS[\phi]}$ }\,,
\end{equation}
where the early time boundary condition is implicitly defined with an $i\epsilon$ prescription and the path integral is done over all field configurations that connect to $\varphi(\bfx)$ at time $\eta$. This path integral representation is useful because it helps organize the perturbative computation of the wavefunction coefficients. The calculation of wavefunction coefficients is quite similar to that of scattering amplitudes (for details see e.g.~\cite{Anninos:2014lwa,Goon:2018fyu,Baumann:2020dch,Goodhew:2020hob}). The primary difference is that there are now two kinds of propagators: a bulk-to-boundary propagator $K(\bf k,\eta)$ that connects vertices to the late-time boundary, and a bulk-to-bulk propagator $G_{\bf k}(\eta_1,\eta_2)$ that connects bulk vertices and which vanishes when one of the times is taken to the boundary. The other main differences is that we only Fourier transform in the spatial directions and integrate over all possible insertion times for interaction vertices. It is these time integrals that are actually the main source of computational difficulty in perturbative bulk calculations.

\vskip6pt
\noindent
{\bf Boundary correlators}\\[2pt]
\noindent
Given the wavefunction~\eqref{eq:wfcoeff}, we obtain equal-time correlators via the usual quantum mechanics procedure of squaring and integrating
\begin{equation}
\langle\varphi(\bfx_1)\cdots \varphi(\bfx_n)\rangle =\frac{ \displaystyle\int{\cal D} \varphi\,\varphi(\bfx_1)\cdots \varphi(\bfx_n) \left\lvert\Psi[\varphi]\right\rvert^2}{\displaystyle\int{\cal D} \varphi \left\lvert\Psi[\varphi]\right\rvert^2}\,.
\end{equation}
If one's interest is directly in these {\it in-in} correlation functions, the detour we have taken through the wavefunction is not necessary. Instead one can apply the so-called {\it in-in} formalism to directly construct these correlators (see, e.g.~\cite{Schwinger:1960qe,Keldysh:1964ud,Weinberg:2005vy,Giddings:2010ui}). This formalism introduces several complications compared to conventional {\it in-out} quantum field theory. Operators are ordered along a multi-branch contour that time-evolves the in vacuum forward in time to the moment of interest for computing the correlation function (the $+$ branch) and then reverse time-evolves back into the infinite past (the $-$ branch). There are then four different propagators $G_{\pm {\hat \pm}}\left(\eta,{\bar \eta}\right)$ used to connect operators inserted on the different branches, where $\pm$ and ${\hat \pm}$ refer, respectively, to the branch of $\eta$ and ${\bar \eta}$.\footnote{This formalism can be implemented via a path integral with doubled field content~\cite{Feynman:1963fq}.} 
Given these propagators, one can define Feynman rules as usual to compute {\it in-in} correlation functions directly.

\subsubsection{Symmetries}

A recurring theme in modern physics is that some of the deepest and most structural insights that we can gain into systems are consequences of symmetry. It is therefore natural to examine the symmetries of the early universe. 
The correlations generated by inflationary dynamics must reflect these underlying symmetries and so they can be used to construct and constrain inflationary observables.

\vskip 4pt
Observations suggest that the background spacetime during the inflationary epoch was very close to being de Sitter space:
\begin{equation}
{\rm d}s^2 = \frac{1}{H^2\eta^2}\left(-{\rm d}\eta^2 +{\rm d}\bfx^2\right) ,
\label{equ:dS-Metric}
\end{equation}
which is a maximally symmetric space with the following Killing vectors:
\begin{equation}
\begin{aligned}
P_i&= \partial_i\,, &  \qquad\quad D & = -\eta \partial_\eta - x^i\partial_i\,,\\
J_{ij} &= x_i\partial_j - x_j\partial_i\,, & K_i &=  2x_i \eta\partial_\eta +\left(2x^jx_i+(\eta^2- x^2) \delta^j_i\right)\partial_j\,.
\end{aligned}
\label{eq:dssymms}
\end{equation}
The symmetries generated by $P_i$ and $J_{ij}$ are the familiar translational and rotational symmetries of the ${\mathbb R}^d$ spatial slices. In addition, de Sitter space possesses a dilation symmetry, generated by $D$ and $d$ boost-like symmetries generated by $K_i$, which act like special conformal transformations on the boundary.
All together, the algebra of isometries for $(d+1)$-dimensional de Sitter space ${\rm dS}_{d+1}$ is $so(d+1,1)$, which coincides with the conformal algebra of $d$-dimensional Euclidean space~$\mathbb{R}^d$.  

\vskip4pt
Fields in de Sitter space behave especially simply at late times. For example, a scalar field, $\phi$, of mass $m$ scales as
\begin{equation}
\phi(\bfx,\eta\to 0)  = {\cal O}_+(\bfx) \eta^{\Delta_+}+{\cal O}_-(\bfx)\eta^{\Delta_-}\,,
\label{eq:massops}
\end{equation}
where its two fall-offs are fixed in terms of its mass as
\begin{equation}\label{mass}
\Delta_{\pm} = \frac{d}{2}\pm i\mu\,\quad{\rm where}\quad i\mu=\sqrt{\frac{d^2}{4}-\frac{m^2}{H^2}}\,,
\end{equation}
which is the analytic continuation of the familiar AdS/CFT relation. From~\eqref{eq:massops}, we infer that the coefficients ${\cal O}_\pm$ transform kinematically like primary operators of weight $\Delta_\pm$ under the action of the conformal group.\footnote{This fact lies at the heart of the proposed dS/CFT correspondence~\cite{Strominger:2001pn,Bousso:2001mw,Strominger:2001gp,Klemm:2001ea,Balasubramanian:2002zh,Maldacena:2002vr,Harlow:2011ke,Anninos:2011ui,Anninos:2012ft,Anninos:2012qw,Anninos:2017eib}. Our discussion does not rely on any detailed microscopic correspondence per se, rather the constraints we discuss are kinematic in nature.} This leads to substantial constraints on the structure of correlation functions of the field $\phi$ in exact de Sitter space. In order to diagonalize the action of translations, it is natural to Fourier transform and treat correlators in momentum space, which if is often done in cosmology. In de Sitter, it is further natural to also diagonalize dilations by going to {\it Mellin space}~\cite{Sleight:2019mgd}. This simplifies many aspects of perturbative bulk calculations, and can be thought of as an analogue of a Fourier transform in the temporal direction, putting the bulk and boundary on somewhat of an equal footing by passing to harmonic space.\footnote{In the same way that spatial integrals of plane waves generate delta functions, Mellin space simplifies integrals over conformal time, which is often the most challenging aspect of bulk perturbation theory. Depending on the physics of interest, many properties of the final correlator can be read off directly in Mellin space, without having to perform the inverse Mellin transform~\cite{Sleight:2019mgd,Sleight:2019hfp}.}

\vskip 4pt
Since de Sitter space is maximally symmetric it does {\it not} describe an evolving universe. The energy density is constant and all spatial slices are equivalent. An important aspect of any inflationary model is therefore the breaking of the de Sitter symmetries, which lead to the Goldstone mode, $\pi(\eta,\bfx)$, and the associated curvature perturbation, $\zeta = - H \pi$~\cite{Creminelli:2006xe,Cheung:2007st}. 
All current observations are reproduced by a nearly scale-invariant two-point function for $\zeta$, and various upper bonds exist on higher-point functions. 
This suggests that the correlations of $\zeta$ is invariant under spatial rotations and translations, and approximately invariant under scale transformations---which can be realized as a diagonal combination of an internal shift symmetry and the (nonlinearly realized) dilation transformation generated by $D$.\footnote{Strictly speaking, none of these assumptions is compulsory given current observations. See~\cite{Endlich:2012pz,Bartolo:2013msa,Kang:2015uha,Slosar:2019gvt} for examples where rotations, translations and/or dilations are broken. Typically these models have internal symmetries that compensate the breaking by some diagonal transformation.}

\vskip 4pt
On the other hand, we have no evidence whether or not de Sitter boosts (generated by $K_i$) are good approximate symmetries of cosmological correlators because the other symmetries already completely fix the only observable we have so far, namely the two-point function.
In the language of cosmological model building, we can say that while the breaking of scale-invariance must be slow-roll suppressed, in principle de Sitter boosts can be arbitrarily badly broken. Depending on the physics of interest, we can therefore proceed  via two paths. We can either assume that de Sitter boosts are also only weakly broken and leverage the full power of de Sitter symmetries, or we can can consider the case where de Sitter boosts are not even approximate linear symmetries (and instead are nonlinearly realized~\cite{Creminelli:2012ed,Assassi:2012zq,Hinterbichler:2012nm,Hinterbichler:2013dpa}), which is the regime where phenomenological signals are larger. Both situations  are of interest and have complementary strengths. It is worthwhile to summarize in a bit more detail these synergies:
\begin{itemize}
    \item \textbf{De Sitter bootstrap:} In situations where both the background and the dynamics are approximately invariant under the full de Sitter group (including slow-roll inflation), we can leverage these symmetries in several ways. Firstly, we can organize observables into representations of the de Sitter group, which highly constrains their form and properties. In addition, we can utilize the isomorphism between the de Sitter group and the Euclidean conformal group to import insights from the study of conformal field theory and of perturbation theory in Anti-de Sitter space into the cosmological setting~\cite{Sleight:2019hfp,Sleight:2020obc,Sleight:2021plv,DiPietro:2021sjt,Hogervorst:2021uvp,Ghosh:2014kba,Albayrak:2018tam,Albayrak:2019asr,Albayrak:2019yve,Albayrak:2020isk,Albayrak:2020bso} (see Section~\ref{sec:AdS}). It is worth noting that there are several cases of phenomenological interest where full de Sitter symmetries are approximate symmetries of the dynamics. In addition to correlators of the inflaton in slow-roll inflation (where, for example, three point functions can be obtained from a deformation of a de Sitter-invariant four-point function~\cite{Kundu:2014gxa,Arkani-Hamed:2015bza,Arkani-Hamed:2018kmz,Baumann:2019oyu}), correlation functions of spectator fields (including the graviton) are invariant under de Sitter symmetries at leading order in slow-roll~\cite{Antoniadis:1996dj,Maldacena:2011nz,Creminelli:2011mw,Antoniadis:2011ib}. Beyond these phenomenological motivations, this highly symmetric situation enables computations that otherwise would be intractable. From these computations, we can abstract lessons that can be applied to more phenomenologically relevant setups.\footnote{Beyond cosmological motivations, there are natural connections to the study of CFTs in momentum space, a subject where there has been much recent activity~\cite{Coriano:2013jba, Coriano:2019nkw, Maglio:2019grh, Bzowski:2013sza, Bzowski:2019kwd, Isono:2018rrb, Isono:2019wex, Bautista:2019qxj, Gillioz:2019lgs, Gillioz:2019iye,Gillioz:2020mdd,Jain:2020rmw,Bzowski:2020kfw,Caron-Huot:2021kjy,Jain:2021vrv,Gillioz:2021kps}.}

    \item \textbf{Boostless bootstrap:} In order to generate large enough interactions to be phenomenologically interesting (at least in the single-field context) it is typically necessary for de Sitter boosts to be strongly violated by interactions of the scalar fluctuations. This is both interesting---because these interactions are less constrained and so more signals are possible---and a challenge, because the symmetries of the problem are reduced. However, there are still many general constraints on the structure of cosmological correlation functions that can be leveraged to bootstrap observables in this setting. (In many cases first seen in more symmetric situations and then abstracted.) For example, features of the singularity structure, or consequences of unitarity continue to hold in these less symmetric settings and can be applied to construct correlators of the inflaton in these models~\cite{Pajer:2020wxk,Jazayeri:2021fvk,Bonifacio:2021azc,Cabass:2021fnw,Meltzer:2021zin}. One advantage of this approach is that the motivations are primarily phenomenological, so one can exploit additional constraints satisfied by the massless particles of interest~\cite{Benincasa:2019vqr,Jazayeri:2021fvk,Baumann:2021fxj,Hillman:2021bnk}. Utilizing these insights, it is possible to compute predictions that capture most realistic models, including single-field models with sizable interactions (non-Gaussianities) in the scalar and tensor sector.
\end{itemize}

\subsubsection{Singularities}
\label{sec:sing}

An inspirational insight from the study of flat-space scattering amplitudes is that the S-matrix is often either completely or mostly fixed by its singularities~\cite{Elvang:2013cua}. For example, the tree level S-matrix can only have pole-like singularities when intermediate particles go on-shell, and the residues of these singularities are products of lower-point amplitudes with positive coefficients (as mandated by unitarity). In many cases this is enough information to completely reconstruct the entire amplitude and can be systematized through powerful recursion relations~\cite{Britto:2005fq,Cheung:2016drk,Elvang:2018dco}. Another inspirational success story is provided by generalized unitarity~\cite{Bern:2011qt}, where at one-loop the discontinuities of amplitudes are expressible in terms of tree-level information and in many cases serves to uniquely specify the full answer.

\vskip 4pt
Our understanding of the properties of the wavefunction and correlators is comparatively more primitive, but much recent progress has been made in the study of their singularities~\cite{Arkani-Hamed:2017fdk,Arkani-Hamed:2018kmz,Arkani-Hamed:2018bjr,Benincasa:2018ssx,Baumann:2020dch,Baumann:2021fxj}. Much as in the case of scattering amplitudes, this information is in many cases sufficient to uniquely reconstruct the entire correlator~\cite{Benincasa:2018ssx,Baumann:2020dch,Baumann:2021fxj}. In other cases additional information must be supplied. This can be done systematically, and so singularities serve as useful anchors where the properties of correlators are known.

\vskip 4pt
We now describe the (tree-level) singularities of the cosmological wavefunction. The wavefunction is somewhat simpler than correlation functions, so its singularities are easier to characterize. There is no loss of generality because the singularities of correlators can be inferred from this information, at least in perturbation theory (see e.g.~\cite{Goodhew:2020hob}).
The singularities of the tree-level wavefunction naturally occur at certain locations in {\it energy} space, where by energies we mean the magnitudes $|\v{k}|$ of the three-momenta that wavefunction coefficients depend on.\footnote{This is a slight abuse of terminology, as energy is not strictly  well-defined in cosmology. Nevertheless, the momentum magnitudes play a similar role in mode functions that energies do in flat space.} 

\vskip 4pt
A ubiquitous singularity occurs when the total energy involved in the process vanishes, $E \equiv \sum k_n \to 0$.  The physical origin of this singularity is the unbounded region of time integration that gives the wavefunction coefficient. Normally, this infinite domain is regulated by oscillatory factors, which vanish exactly when these sums of energies vanish, so the wavefunction diverges.  The coefficient of this {\it total energy singularity} is the corresponding flat-space scattering amplitude~\cite{Maldacena:2011nz,Raju:2012zr}
\begin{equation}\label{kTeq0}
\lim_{E\to 0} \psi_n = \frac{iA_n}{E^\alpha}\,.
\end{equation}
This fact---that cosmological correlation functions (or the wavefunction) have within them a singularity whose residue is a scattering amplitude---provides a beautiful connection between the study of cosmological correlators and that of the S-matrix. The precise order of the singularity, $\alpha$, and even its nature depends on the details of fields and interactions, but the existence of some singularity is robust.\footnote{In flat space, all singularities are simple poles, but in cosmological backgrounds both higher-order poles and branch point singularities can occur, depending on the situation.} 

\vskip 4pt
More generally, wavefunction coefficients have singularities whenever the energies $E_\gamma$ flowing into a connected subdiagram $\gamma$ vanish. These {\it partial energy singularities} are a characteristic feature of particle exchange.
The residues of these singularities are related to lower-point scattering amplitudes and wavefunction coefficients. For example, for an $n$-point wavefunction coefficient, the singularity where the partial energy $E_\gamma$ vanishes (splitting the graph into two sub-graphs $\gamma$ and $\gamma'$) is of the following schematic form:
\begin{equation}
\lim_{E_\gamma\to 0}\psi_{n} = \frac{iA_\gamma \times \widetilde\psi_{\gamma'}}{E_{\gamma}^\beta}\,.
\label{equ:partial}
\end{equation}
Here $\widetilde\psi_{\gamma'}$ is a 
shifted version of the ($m$-point) wavefunction associated to the subgraph $\gamma'$:
\begin{equation}
\widetilde\psi_{\gamma'}(k_{1},\cdots, k_{m}, k_I) \equiv P(k_I)\Big(\psi_{\gamma'}(k_{1},\cdots, k_{m}, -k_I)-\psi_{\gamma'}(k_{1},\cdots, k_{m}, k_I)\Big)\,,
\end{equation}
where the factor $P(k_I)$ is the power spectrum of the exchange field connecting the two graphs, a structure whose origin will become clear in Section \ref{sec:COT}.  In flat space, the singularities are always simple poles, $1/E_\gamma$. Conversely, in general FLRW spacetime and in dS higher-order poles arise. The order $\beta$ of the pole in (\ref{equ:partial}) can be fixed using dimensional analysis and scale invariance. For example, for the  case of massless scalars and gravitons, it is given by~\cite{Pajer:2020wxk,Hillman:2021bnk}
\begin{equation}
\label{psumA}
\beta\leq 1+\sum_{V\in \gamma} \Big[\text{dim}_V -(d+1)\Big]\,,
\end{equation}
where the sum runs over all vertices $V$ in the (sub)diagram and $\text{dim}_V$ denotes the mass dimension of the vertex $V$. The inequality allows for the possibility that the residues of the highest-order pole vanishes, which can happen in various interesting cases because of symmetry \cite{Pajer:2020wxk,Grall:2020ibl}. Finally, when $\beta$ vanishes there can be branch point singularities whose coefficients are fixed by the cosmological optical theorem (see Section~\ref{sec:Unitarity}).

\begin{figure}[t!]
    \centering
    \includegraphics[width=0.85\textwidth]{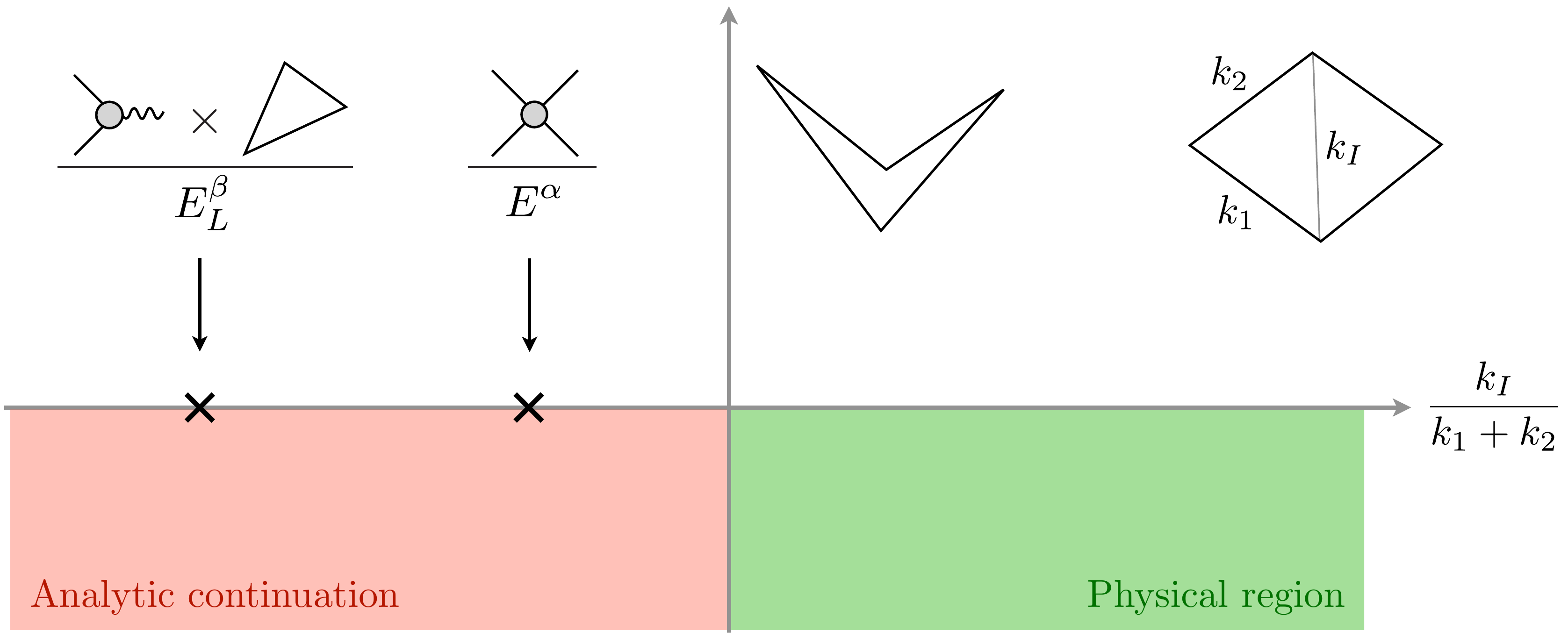}
    \caption{Schematic of the singularities of the cosmological wavefunction. The wavefunction has singularities when (partial) energies are conserved. The coefficients of these singularities can be written in terms of scattering amplitudes and lower-point wavefunctions. Even though these singularities lie at analytically continued momentum configurations, they nevertheless partially control the behavior of the wavefunction for physical kinematics. Extending away from these singularities to general momentum configurations is a boundary version of the challenge of time evolution in the bulk.
    }
    \label{fig:Singularities}
\end{figure}

\vskip 4pt
Importantly, none of these singularities are {\it physical}, in the sense that they cannot be probed by any physical process with real momenta. Nevertheless, they can be accessed by extending momenta into the complex plane, and remarkably, they control the form of correlation functions even in the physical region (see Fig.~\ref{fig:Singularities}). Indeed, one way to phrase the challenge of constructing the boundary wavefunction is: given knowledge of its form in the vicinity of its singularities, how do we extend away from these special loci to general kinematics (in particular into the physical region).
We now review several approaches to meet this challenge. 


\subsubsection{Manifest Locality}\label{sec:MLT}

Locality already played a role in the previous discussion by ensuring that the only poles that appear are at vanishing partial energies. However, in the special case of massless scalars and spin-2 fields (gravitons), there is another manifestation of locality, which provides a powerful bootstrap tool. The key observation is that the massless de Sitter mode function, $K(\eta,k)$, corresponding to the conformal dimension $\Delta=d$, and all of its time derivatives, satisfy
\begin{equation} \label{BulkMLT}
\dfrac{\partial}{\partial k} \left( \frac{{\rm d}^N}{{\rm d}\eta^N} K(\eta,k) \right) \bigg\rvert_{k=0}=0\,.
\end{equation}
If interactions are manifestly local, meaning that they are products of fields and their derivatives at the same spacetime point, this property is inherited by the wavefunction coefficients in perturbation theory and goes under the name of Manifestly Local Test (MLT) \cite{Jazayeri:2021fvk}
\begin{equation} \label{MLTGeneral}
\dfrac{\partial }{\partial k_{c}}\psi_{n}(k_1,\cdots,k_n;\{p\};  \{\bfk\})\Big|_{k_{c}=0}=0\,,
\end{equation}
where the notation indicates that the derivative with respect of an external energy $k_c$ should be taken while keeping fix all energies $\{p\}$ running along internal lines and all vector contractions $\v{k}_a\cdot \v{k}_b$, if present. The result in \eqref{MLTGeneral} can be equivalently derived demanding the absence of non-partial energy singularities. As we review in Section \ref{BBBB}, with this simple condition one can derive all boost-breaking tree-level three- and four-point functions  \cite{Jazayeri:2021fvk,Bonifacio:2021azc}, which are the main target of cosmological observations. Furthermore, the MLT can also be used in combination with partial-energy recursion relations (see Section \ref{PERR}) to fix terms that are not constrained by unitarity. 


\subsection{Bootstrapping Inflationary Correlators}
\label{subsec::BIC}
In many cases the structure of tree-level wavefunctions and correlators is sufficiently rigid that they can be completely reconstructed from information about their singularities and symmetries~\cite{Arkani-Hamed:2017fdk,Arkani-Hamed:2018kmz,Benincasa:2018ssx,Benincasa:2019vqr,Baumann:2019oyu,Baumann:2020dch,Pajer:2020wxk,Jazayeri:2021fvk,Bonifacio:2021azc,Melville:2021lst,Baumann:2021fxj,Bittermann:2022nfh}. In this section, we outline the logic and present some success stories of this reasoning. We first describe how the symmetries of de Sitter space can be used to control the structure of perturbative cosmological correlation functions, which captures the physics of slow-roll inflation. We then review how boost-breaking interactions can also be computed from their singularities, locality and unitarity.


\subsubsection{De Sitter Four-Point Functions}
When the de Sitter symmetry is only softly broken by interactions, the correlation functions of both spectator fields and the inflaton must be compatible with the symmetries~\eqref{eq:dssymms}. This places strong constraints on their form (which are essentially the same kinematic requirements that constrain correlation functions in a conformal field theory). Translation and rotation invariance are easy to satisfy in momentum space: correlation functions that transform covariantly under rotations and which satisfy momentum conservation are compatible with these constraints. The nontrivial consequence of de Sitter symmetry are therefore the {\it kinematic conformal Ward identities} associated to dilations and de Sitter boosts:
\begin{equation}
 \begin{aligned}
 \left[ \hskip 2pt-d+ \sum_{a=1}^n D_{a}  \right] \langle {\cal O}_1 \cdots {\cal O}_a \cdots {\cal O}_n \rangle &=0\, ,   \\[4pt]
 \sum_{a=1}^n K_{a}^i\, \langle {\cal O}_1 \cdots {\cal O}_a \cdots {\cal O}_n \rangle &=0
  \, ,
\end{aligned}
\label{equ:WardID}
\end{equation}
where we have introduced the shorthand notation ${\cal O}_a \equiv {\cal O}_{\bfk_a}$. The operators $D_a$ and $K^i_a$ in~\eqref{equ:WardID} are the Fourier space representation of the dilation and de Sitter boost generators~\eqref{eq:dssymms}, so that the correlators obey differential equations in the momentum variables~\cite{Maldacena:2011nz,Creminelli:2012ed,Bzowski:2013sza,Hinterbichler:2013dpa,Arkani-Hamed:2018kmz,Baumann:2020dch}. (The equations satisfied by the wavefunction coefficients are essentially identical.) Note that these differential equations are the same as those satisfied by correlation functions in a CFT${}_d$ as a consequence of kinematic conformal invariance. This connection allows for an fruitful exchange of ideas between the two subjects. 

\vskip 4pt
Correlators or wavefunction coefficients involving massless particles with spin are further constrained beyond these kinematic requirements. This is essentially because massless particles correspond to conserved currents from the boundary point of view, and their correlations must be compatible with this conservation. These additional constraints are a boundary manifestation of bulk gauge invariance and require that spinning correlators additionally satisfy {\it current conservation Ward--Takahashi (WT) identities}. For example, for a massless spin-1 field these take the form~\cite{Maldacena:2011nz,Bzowski:2013sza,Baumann:2020dch}
\begin{equation}
k_1^i \langle J^i_{\bfk_1} {\cal O}_{\bfk_2} \cdots {\cal O}_{\bfk_n}  \rangle = - \sum_{a=2}^n e_a \langle {\cal O}_{\bfk_2} \cdots {\cal O}_{\bfk_a + \bfk_1} \cdots {\cal O}_{\bfk_n}  \rangle\,,
\label{eq:spin1WT}
\end{equation}
where $e_a$ are the charges of the various operators appearing in the correlator. Notice that this fixes the longitudinal part of the correlator in terms of a lower-point function, demonstrating that it is constrained in terms of other data.\footnote{The fact that conserved operators have additional requirements beyond kinematic de Sitter invariance in order for their longitudinal modes to decouple reflects that they are not phrased in terms of completely unconstrained kinematic variables. This suggest that a more elegant treatment of spinning correlation functions in de Sitter space exists, and finding it is an interesting challenge for the future.}

\vskip 4pt
To bootstrap the boundary correlators directly, we must therefore solve the differential equations implied by the conformal Ward identities~\eqref{equ:WardID} (along with the WT identities in situations with spin). 
In order for this task to be tractable, we need some principles to select the physical solutions of interest from the infinite number of solutions to these equations. In this regard singularities play an important role. First, we can restrict the space of possible solutions by restricting to functions that only have the singularities associated to tree-level exchange (for example). Further, to select the physical solution within this class, we forbid the presence of unphysical singularities, and further require that the physical singularities are normalized correctly. In the following, we will give a concrete example of this logic.

\vskip6pt
\noindent
{\bf A simple seed: conformal scalars}\\[2pt]
\noindent
Three-point functions in de Sitter space are highly constrained by the boundary conformal symmetry, being essentially unique up to a finite number of constants depending on the field content~\cite{Polyakov:1970xd,Osborn:1993cr,Maldacena:2011nz,Creminelli:2011mw,Bzowski:2013sza,Pajer:2016ieg}. Consequently, the first nontrivial dynamical information about a theory arises at four points.
As a simple example of the bootstrap procedure, we derive the four-point correlation function of conformally coupled scalars in exact de Sitter space~\cite{Arkani-Hamed:2015bza,Arkani-Hamed:2018kmz,Baumann:2019oyu}. Despite its simplicity, this solution is important, as it can be transformed into correlation functions of more physical interest. For conformal scalars, $\varphi$, the differential equations~\eqref{equ:WardID} can be combined and the rescaled correlator $F \equiv \lvert \bfk_1+\bfk_2\rvert\,\langle\varphi_1\varphi_2\varphi_3\varphi_4\rangle$ satisfies~\cite{Arkani-Hamed:2015bza,Arkani-Hamed:2018kmz}
\begin{equation}
\left(\Delta_u-\Delta_v\right) F(u,v) = 0\,,
\label{eq:genconf4pt}
\end{equation}
where we have introduced the differential operator
\begin{equation}
\Delta_u \equiv u^2(1-u^2)\partial_u^2-2u^3\partial_u\,,
\end{equation}
with $u\equiv \lvert\bfk_1+\bfk_2\rvert/(k_1+k_2)$. The operator $\Delta_v$ is defined similarly in terms of $v\equiv \lvert\bfk_3+\bfk_4\rvert/(k_3+k_4)$.

\vskip 4pt
Any possible bulk process will generate a boundary correlation function that solves the partial differential equation~\eqref{eq:genconf4pt}. Specializing to tree-level particle exchange, however, this equation can be split into 
a  pair of ordinary differential equations~\cite{Arkani-Hamed:2015bza,Arkani-Hamed:2018kmz}
\begin{equation}
\begin{aligned}
\left(\Delta_u+M^2\right)F &=C_n\,, \\ \left(\Delta_v+M^2\right)F &=C_n\,,
\end{aligned}
\label{eq:exceqs}
\end{equation}
where $C_n$ is a solution to~\eqref{eq:genconf4pt} that has {\it only} a total energy singularity and hence corresponds to a contact interaction in the bulk. The simplest contact solution is $C_0 = uv/(u+v)$ and higher-order contact solutions $C_{n>0}$ are obtained by repeated application of $\Delta_u$. It is straightforward to check that any solution to the pair of equations in \eqref{eq:exceqs} will also solve~\eqref{eq:genconf4pt}. However, the structure of these ODEs restricts the singularity structure of solutions to be that arising from bulk tree exchange, where $M^2$ is set by the mass of the exchanged particle. These are second-order differential equations, so we require two boundary conditions to solve them. A generic solution will not only have the expected total and partial energy singularities---at $u \to - v$ and $u \to -1$, respectively---but will also have a {\it folded singularity} at $u \to + 1$.
The latter corresponds to a momentum configuration where the quadrilateral formed by the momenta in Fourier space degenerates to a triangle because two of the momentum becomes colinear (here, $\bfk_1$ and $\bfk_2$).
In Bunch--Davies initial conditions, folded singularities are unphysical and should therefore be absent~\cite{Arkani-Hamed:2018kmz,Green:2020whw}. 
One boundary condition is therefore to forbid the presence of folded singularities.  
A second boundary condition is provided by normalizing one of the physical singularities correctly. The other physical singularities then become consistency checks.

\vskip 4pt
In~\cite{Arkani-Hamed:2018kmz}, the equations~\eqref{eq:exceqs} were solved for arbitrary values of $M^2$. The explicit form of the solution isn't very illuminating and therefore won't be displayed here. Let us just remark that the solution has a piece corresponding to the EFT expansion of the bulk interactions (arising from integrating out the massive particles) and a piece describing the production and decay of massive particles in the expanding spacetime. 
The latter is a consequence of the time-dependence of the background, which is reproduced without any explicit reference to the bulk time evolution. In this sense, the bootstrap procedure has realized the goal of describing ``time without time".

\vskip 4pt
To illustrate the physical principles that select the solution more explicitly, it is useful to consider the particular case where $M^2 = 0$. This corresponds to conformally coupled scalars exchanging a conformal scalar in the bulk. In this case, the most general solution to~\eqref{eq:exceqs} with $C_0$ as a source is
\begin{equation}
\begin{aligned}
F = \,&\frac{1}{2}{\rm Li}_2\left(\frac{u(1-v)}{u+v}\right)+\frac{1}{2}{\rm Li}_2\left(\frac{v(1-u)}{u+v}\right)+\frac{1}{2}\log\left(\frac{u(1+v)}{u+v}\right)\log\left(\frac{v(1+u)}{u+v}\right)+\frac{\pi^2}{6}\\
&+c_1+c_2\log\left(\frac{1-u}{1+u}\right)\log\left(\frac{1-v}{1+v}\right)\,,
\end{aligned}
\label{eq:cccccex}
\end{equation}
where ${\rm Li}_2(x)$ is the dilogarithm and $c_1$ and $c_2$ are arbitrary constants. We now need to impose boundary conditions: 
the absence of folded singularities requires us to set $c_2=0$ and the normalization of the partial energy singularity $u\to -1$ sets $c_1=0$, leaving the first line of~\eqref{eq:cccccex} as the physical solution~\cite{Arkani-Hamed:2015bza,Arkani-Hamed:2018kmz}.\footnote{Note that the four-point wavefunction and correlator differ by a constant factor, accounting for the difference in the factorization limits of the two objects. See~\cite{Hillman:2019wgh} for an explicit expression for the wavefunction.}

\vskip6pt
\noindent
{\bf Massive exchanges and weight-shifting}\\[2pt]
\noindent
The equations~\eqref{eq:exceqs} can be solved for arbitrary values of $M^2$, which produces the boundary correlation function for four conformally coupled scalars that arises from the exchange of a massive scalar. Though the detailed form is complex,   it admits a rapidly-convergent power series representation so for practical purposes it is very efficient~\cite{Arkani-Hamed:2018kmz}. We will denote this solution abstractly as $F_{(M)}$. The situation that we have considered may seem somewhat artificial and unphysical, but remarkably the solution $F_{(M)}$ can be efficiently transformed into other solutions of interest by utilizing so-called {\it weight-shifting} operators~\cite{Baumann:2019oyu} which were first constructed in the study of conformal field theory~\cite{Costa:2011dw,Karateev:2017jgd}. Conceptually, these operators map solutions of the kinematic Ward identities~\eqref{equ:WardID} to new solutions with different values of masses and spins for both external and internal operators. This allows us to generate a wide menu of physical processes from $F_{(M)}$, which acts as a simple seed object~\cite{Arkani-Hamed:2018kmz,Baumann:2019oyu,Baumann:2020dch}. This technique is also readily applied to situations beyond the four-point function. The fact that all of these different physical solutions can be mapped to each other by differential operations reveals a unity of perturbative de Sitter physics that is completely invisible in the standard Lagrangian approaches.

\begin{figure}[t]
    \centering
    \includegraphics{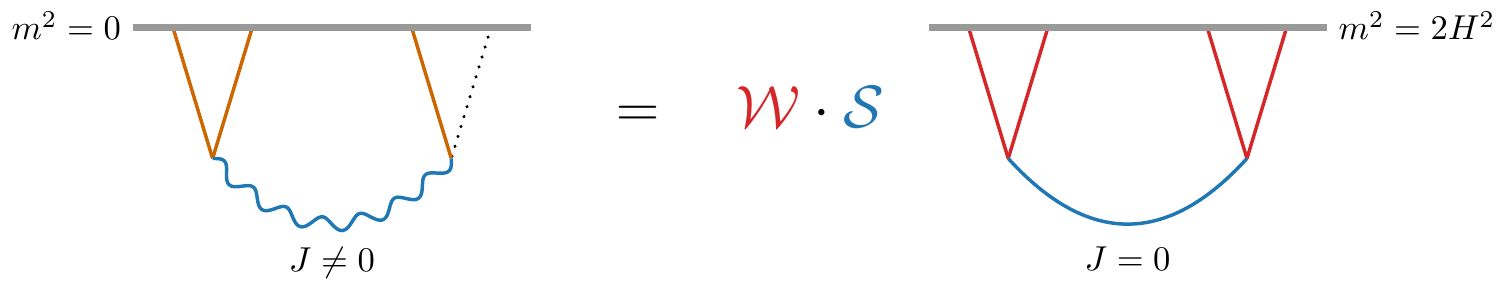}
    \caption{The inflationary three-point function due to the exchange of a massive spinning particle inherits the analytic structure of the four-point function of conformally coupled scalars exchanging a massive scalar. To obtain the inflationary correlator, we apply a spin-raising operator ${\cal S}$ (changing the spin of the exchanged particle) and a weight-shifting operator~${\cal W}$ (converting the external particles from conformally coupled to massless). Finally, taking one of the legs to be soft gives the inflationary three-point function in the slow-roll limit. }
    \label{fig:weightshifting}
\end{figure}

\vskip6pt
\noindent
{\bf Inflationary three-point functions}\\[2pt]
\noindent
In the slow-roll limit, where de Sitter symmetries are softly broken by the interactions of the inflaton, the three-point correlator of $\zeta$ can be obtained as a deformation of an exactly de Sitter invariant four-point function, evaluated in the soft limit~\cite{Kundu:2014gxa,Arkani-Hamed:2015bza,Arkani-Hamed:2018kmz,Baumann:2019oyu}. Combined with the weight-shifting procedure, this makes it possible to efficiently generate inflationary three-point functions arising from the exchange of particles of arbitrary mass and spin~\cite{Arkani-Hamed:2018kmz,Baumann:2019oyu}
\begin{equation}
\langle \zeta_{\bfk_1}\zeta_{\bfk_2}\zeta_{\bfk_3}\rangle^{(M,J)} = -\epsilon\, k_3^3\, P_J(\alpha)\, U_{12}^{(J,0)} F_{(M)}^{(J)}(u,1)\, +\, {\rm perms}. \, ,
\label{equ:InfB}
\end{equation}
where ``perms" denotes symmetrization over the external momenta,
$P_J(\alpha)$ is a Legendre polynomial with $\alpha\equiv (k_1-k_2)/k_3$, and $J$ denotes the spin of the exchanged particle. 
Here, $F_{(M)}^{(J)}$ is the solution to the conformal kinematic Ward identities for conformal scalars exchanging a particle of mass $M$ and spin-$J$, which can be obtained by weight-shifting the scalar solution $F_{(M)}$. The operator $U_{12}^{(J,0)}$ is another weight-shifting operator that changes the mass of the external scalar field to be that of the inflaton $\zeta$. Notice that this bispectrum is proportional to $\epsilon\equiv -\dot H/H^2$, which is the slow-roll parameter that measures the deviation of the background from de Sitter space.
From a phenomenological point of view, these bispectra are interesting because they display a characteristic oscillatory feature in the squeezed limit (where $\bfk_1\ll \bfk_2, \bfk_3$) that scales as~\cite{Chen:2009zp,Arkani-Hamed:2015bza}
\begin{equation}
\lim_{\bfk_1\to0}\langle \zeta_{\bfk_1}\zeta_{\bfk_2}\zeta_{\bfk_3}\rangle^{(M,J)} \sim \left(\frac{k_1}{k_2}\right)^{\frac{3}{2}+i\mu}+{\rm c.c.}\,,
\label{eq:massscalesqueezed}
\end{equation}
where the exponent is set by the mass of the exchanged field via $i\mu \equiv \sqrt{d{}^2/4-M^2/H^2}$~\cite{Chen:2009zp,Baumann:2011nk,Noumi:2012vr,Arkani-Hamed:2015bza}. In addition, the angular structure of the correlator is determined by the spin of the exchanged particle~\cite{Arkani-Hamed:2015bza}. In principle, these characteristic features of particle exchange would allow us to do ``cosmic spectroscopy" and use the inflationary epoch as a sort of particle collider if these features are present. This has motivated a large measure of interest in the area of {\it cosmological collider physics}~\cite{Chen:2009zp, Baumann:2011nk, Assassi:2012zq, Chen:2012ge, Pi:2012gf, Noumi:2012vr, Baumann:2012bc, Assassi:2013gxa, Gong:2013sma, Arkani-Hamed:2015bza, Lee:2016vti, Kehagias:2017cym, Kumar:2017ecc, An:2017hlx, An:2017rwo, Baumann:2017jvh, Kumar:2018jxz, Bordin:2018pca, Goon:2018fyu, Anninos:2019nib, Kim:2019wjo, Alexander:2019vtb, Hook:2019zxa, Kumar:2019ebj, Liu:2019fag, Wang:2019gbi, Chen:2018uul, Chen:2018brw, Lu:2021wxu, Wang:2021qez, Pinol:2021aun, Cui:2021iie}.

\vskip6pt
\noindent
{\bf Spinning correlators}\\[2pt]
\noindent
All inflationary models have two compulsory light degrees of freedom: the Goldstone of spontaneously broken time translations, $\pi$, and the graviton, $\gamma_{ij}$. It is therefore important to understand the properties of correlation functions involving particles with spin in inflationary spacetimes.
In addition to the kinematic constraints of conformal symmetry (which are substantially more complex in the spinning case), correlation functions with massless fields with spin must satisfy WT identities~\eqref{eq:spin1WT}. This means that massless spinning fields are more constrained than their scalar counterparts. This is to be expected, indeed it is well known in the scattering context that space of allowed interactions is extremely limited, with theories like Yang--Mills and General Relativity being essentially unique consistent theories. Similarly, in the cosmological context, it is not always possible to simultaneously solve the constraints of kinematic conformal invariance and current conservation~\cite{Baumann:2020dch,Sleight:2021iix}. Indeed, requiring that correlation functions solve both equations at the same time reproduces, from the boundary perspective, the relations between couplings required by consistent gauge theories~\cite{Baumann:2020dch,Sleight:2021iix}. 
Moving forward, it will be interesting to fully map out the space of consistent theories involving massless particles in cosmology. For instance, there exist interesting types of gauge fields in de Sitter space (partially massless particles) that do not have any direct flat-space counterparts~\cite{Deser:1983mm,Brink:2000ag,Deser:2001us}. Progress has been made in the study of their correlation functions in the presence of matter~\cite{Sleight:2021iix}, ruling out certain matter couplings, and an interesting question is whether a theory with a finite number of such particles can be consistent (given the existence of some no-go results~\cite{Deser:2012qg,deRham:2013wv,Joung:2014aba,Garcia-Saenz:2015mqi,Joung:2019wwf}). Such a question can reasonably be answered from the bootstrap perspective.

\subsubsection{Boostless Three-Point Functions}\label{BBBB}

To generate any sizable non-Gaussian $n$-point function in single-field inflation requires that the de Sitter boost symmetry is strongly broken by the interactions of $\zeta$ \cite{Green:2020ebl}. It is highly desirable, therefore, to be able to apply the bootstrap methodology to compute correlation functions in these less symmetric theories. Since we no longer have the constraints of de Sitter boosts, one might expect that this situation is too unconstrained to make progress. However, the singularities of consistent correlators along with information about locality and the remaining symmetries can still be used to reconstruct the boundary correlators~\cite{Pajer:2020wxk,Grall:2020ibl,Meltzer:2021zin,Jazayeri:2021fvk,Bonifacio:2021azc,Cabass:2021fnw,Baumann:2021fxj,Hillman:2021bnk}. It is somewhat remarkable that inflationary correlators can be bootstrapped with such minimal input. In the following, we briefly review some success stories of this approach.

\vskip6pt
\noindent
{\bf Bispectrum of the EFT of inflation}\\[2pt]
\noindent
We can construct the bispectrum in the EFT of inflation by first writing down the most generic ansatz compatible with a small set of ``boostless bootstrap rules" that enforce symmetries and basic physical principles such as locality and then using the MLT (discussed in Section \ref{sec:MLT}) to fix all remaining coefficients \cite{Pajer:2020wxk,Jazayeri:2021fvk}. The result is the most general tree-level bispectrum to all orders in the derivative expansion. For simplicity, we discuss here only scalars, but a similar approach has been used to bootstrap also all graviton bispectra \cite{Cabass:2021fnw}. The bootstrap rules are: (i) invariance under translations rotations and dilations (but no assumption about boosts) fixes the kinematic variables, (ii) massless fields at tree-level in dS ensures that the result is a rational function, (iii) Bose symmetry enforces permutation invariance and finally (iv) locality and the choice of the Bunch--Davies vacuum imply that the only singularity is at vanishing total energy, $k_T\equiv k_1+k_2+k_3$. Under these assumptions, the most general bispectrum must take the form~\cite{Pajer:2020wxk,Jazayeri:2021fvk}
\begin{equation}
\langle \zeta_{\bfk_1}\zeta_{\bfk_2}\zeta_{\bfk_3}\rangle = \frac{1}{(k_1k_2k_3)^3}\sum_p \frac{{\rm Poly}_{p+3}(k_1,k_2,k_3)}{k_T^{p}}\,,
\label{eq:EFTinfbispec}
\end{equation}
where ${\rm Poly}_{p+3}(k_1,k_2,k_3)$ is a polynomial with mass dimension $p+3$ that is symmetric in its arguments and can therefore be written uniquely in terms of elementary symmetric polynomials. The order $p$ of the pole is fixed by \eqref{psumA} and is found to coincide with the total number of derivatives in the corresponding interaction. Terms with larger $p$ are higher order in the EFT derivative expansion. The set of all possible polynomials ${\rm Poly}_{p+3}$ is given by all possible solutions of the MLT \cite{Jazayeri:2021fvk}, which requires 
\begin{equation}
\partial_{k_1}\Big(k_1^3\langle \zeta_{\bfk_1}\zeta_{\bfk_2}\zeta_{\bfk_3}\rangle\Big)\bigg\rvert_{k_1 = 0} = 0\,.
\end{equation}
This ensures that the bispectrum arose from a local bulk interaction involving massless inflatons. Taking account of all of these constraints, the number of free parameters in~\eqref{eq:EFTinfbispec} can be shown to exactly match the free parameters in the EFT of inflation Lagrangian, order by order in $p$~\cite{Pajer:2020wxk}. The result satisfies some general properties: the leasing residue of the $k_T\to0$ pole is determined by a corresponding flat-space scattering amplitude, which needs not be invariant under Lorentz boosts \cite{Pajer:2020wnj}, and it satisfies the single-field soft limit that are a consequence of the nonlinearly realized symmetries of $\zeta$~\cite{Creminelli:2012ed,Assassi:2012zq,Hinterbichler:2012nm,Hinterbichler:2013dpa,Pajer:2017hmb,Bordin:2017ozj,Jazayeri:2019nbi,Avis:2019eav}.

\vskip 4pt
A similar derivation to the one above was carried out for the contact four-point function in~\cite{Bonifacio:2021azc}. To describe boostless exchange diagrams, we need unitarity in the form of the cosmological optical theorem to be discussed in Section \ref{sec:COT}.

\vskip6pt
\noindent
{\bf Graviton correlators}\\[2pt]
\noindent
 In addition to the inflationary scalar correlators, techniques that don't rely on de Sitter boost symmetries can also be applied in the construction of graviton correlators. In the simplest and most studied inflationary setup of the EFT of inflation, graviton interactions are de Sitter invariant to leading order in the slow-roll expansion. Hence, there are only a finite number of possibilities at three points (three possible wavefunction coefficients producing only {\it two} cubic correlators in $D=4$)~\cite{Maldacena:2011nz}. However, in more general models of inflation such as Solid Inflation~\cite{Endlich:2012pz,Ricciardone:2016lym,Piazza:2017bsd}, gravitons can also be sensitive to the departure of the background from pure de Sitter space, leading to a larger menu of possible shapes and non-Gaussianities that can be large enough to be detected by cosmological surveys (see \cite{Creminelli:2014wna,Bordin:2017hal,Bordin:2020eui,Cabass:2021iii} for related discussions). Here, we wish to focus on one interesting feature that arises for parity-odd graviton bispectra at tree-level. Such a signal cannot arise in the presence of de Sitter boost symmetry. Instead, for general boost breaking theories one naively expects infinitely many possible shapes corresponding to the  infinitely many cubic Lagrangian operators with an ever increasing number of derivatives. Remarkably unitarity dictates that only three shapes are allowed, as we will now see.
 
 \vskip 4pt
 We can write any $n$-point correlator from a single contact interaction in terms of the wavefunction coefficients as~\cite{Cabass:2021fnw} 
\begin{align} \label{WFtoCorrelator}
B_{n}^{\text{contact}}(\{k \}; \{\bfk\}) =  -\frac{\psi_{n}(\{k \}; \{\bfk\})+ \psi^{\ast}_{n}(\{k \}; -\{\bfk\})}{\prod_{a=1}^{n} 2 \hskip 2pt \text{Re} ~ \psi_{2}(k_{a})}\,.
\end{align}
The crucial point is that unitarity in the form of the cosmological optical theorem (see Section~\ref{sec:COT}) implies that $ \psi_n(\{ k \};  \{\bfk\}) + \psi^*_n(\{ -k\}; - \{\bfk\} ) = 0$. Hence, if $\psi_n(k)$ happens to be even under flipping the sign of its arguments, then the corresponding contact correlators will vanish. Focusing on the bispectrum and writing down a generic ansatz compatible with the boostless bootstrap rules, analogous to \eqref{eq:EFTinfbispec}, one discovers that only terms without any total energy pole have this property.\footnote{There is an infinite number of possible nonzero wavefunction coefficients, but all but three drop out of the bispectrum. In particular, the single parity-odd interaction compatible with de Sitter boost symmetry in $D=4$ cannot contribute to the bispectrum~\cite{Soda:2011am,Shiraishi:2011st}.} This implies that (i) parity-odd correlators from contact interactions are actually regular at vanishing total energy and (ii) there are only a handful of possibles shapes because the degree of the arbitrary polynomial is fixed by scale invariance. For three gravitons there are only three allowed shaped, for graviton-scalar-scalar also three, and only a single shape for graviton-scalar-scalar. Explicit expressions for these shapes are given in \cite{Cabass:2021fnw} and provide important targets for non-Gaussian searches in the graviton sector in the case of a detection of primordial tensor modes.

\vskip 4pt
This example is a poster child of the bootstrap approach because it displays the stark juxtaposition between the Lagrangian description and the actual observables: there are an infinite number of EFT vertices that cannot be removed by field redefinitions, but nevertheless they do not contribute to the parity-odd graviton three-point function. Conversely, from the bootstrap perspective it is immediate to see that only a handful of possible observables are allowed.

\vskip6pt
\noindent
{\bf Building to higher points}\\[2pt]
\noindent
The techniques discussed in this section are more widely applicable beyond the examples that we have outlined. In particular, these boostless approaches are well-suited to construct correlations/wavefunctions of massless particles, which are typically rational functions. As such, these tools can be adapted to also study correlation functions of massless spin-1 and spin-2 fields in the de Sitter setting~\cite{Goodhew:2020hob,Baumann:2020dch}. In addition, these techniques readily generalize to higher points~\cite{Bonifacio:2021azc}, and can be combined with recursive techniques in order to construct a wide variety of rational correlators~\cite{Jazayeri:2021fvk,Meltzer:2021zin,Baumann:2021fxj,Hillman:2021bnk}.

\subsection{Cosmological Polytopes and Beyond}

The disappearance of bulk time and its encoding in the momentum dependence of boundary observables is inspiring, and it emboldens us to consider the more ambitious task of removing the entire spacetime from the picture, to have it reemerge as a derived concept. Indeed, similar attitudes have been remarkably powerful in the study of scattering amplitudes, where many deep structures have been uncovered by phrasing constructions in auxiliary or dual spaces.

\vskip 4pt
In the cosmological context, one notable development in this direction is the study of {\it cosmological polytopes}. These are geometrical objects whose volumes compute the (rational) wavefunction coefficients of a conformally coupled scalar~\cite{Arkani-Hamed:2017fdk}. These rational wavefunctions can then naturally be integrated to construct the wavefunction of a conformal scalar in more general FLRW spaces. It is interesting and intriguing that an object describing cosmology (the wavefunction) arises from a purely geometric question that does not obviously have anything to do with physics.

\vskip 4pt
Aside from being of intrinsic interest as an example of spacetime physics arising from an auxiliary structure, the study of cosmological polytopes---or more generally of simplified models of the wavefunction---has been of great utility in revealing an underlying simplicity in cosmological perturbative dynamics~\cite{Arkani-Hamed:2017fdk,Arkani-Hamed:2018bjr,Benincasa:2018ssx,Benincasa:2019vqr,Hillman:2019wgh,Benincasa:2020aoj}. In these simplified models it is often possible to compute to high multiplicity and prove general statements about the structure of the theory. For example, these models were instrumental in elucidating the structure of singularities of the cosmological wavefunction, which was then abstracted to more general settings~\cite{Arkani-Hamed:2017fdk,Arkani-Hamed:2018bjr,Benincasa:2018ssx}. Another example involves the transformation of flat-space wavefunctions to their de Sitter or FLRW counterparts~\cite{Benincasa:2019vqr}, which can also be applied to more general theories~\cite{Baumann:2021fxj,Hillman:2021bnk}. Cosmological polytopes have also enabled the construction of powerful recursion relations for conformally coupled scalars~\cite{Arkani-Hamed:2017fdk,Hillman:2019wgh} and have been important in the study of how locality and unitarity manifest at the level of the wavefunction~\cite{Benincasa:2020aoj}.

\vskip 4pt
Looking forward, we expect these investigations to yield further insights. Interesting questions to explore include searching for analogues of these geometric structures for spinning wavefunctions and further elucidating how aspects of unitarity manifest themselves geometrically. An important challenge is to understand how these geometric structures (which are defined diagram-by-diagram in perturbation theory) fit together into a larger structure in theories that require diagrams to be combined in observables, like in gauge theories and theories of Goldstones. This would provide an illuminating step toward moving ``beyond Feynman diagrams" in cosmology.

\section{Unitarity in Cosmology} 
\label{sec:Unitarity}

In this section, we summarize recent progress on deriving the consequences of unitarity for cosmological correlators. The discussion is organized into two parts. First, we discuss a cosmological equivalent of the optical theorem \cite{Goodhew:2020hob} and the associated cosmological cutting rules \cite{Melville:2021lst,Goodhew:2021oqg,Baumann:2021fxj} (see~\cite{Meltzer:2019nbs,Meltzer:2020qbr} for an AdS version). This is a genuinely perturbative result that is valid to all orders in perturbation theory, including any number of loops. It applies to fields of any mass and spin with arbitrary local interaction on any FLRW spacetime (including de Sitter) that admit a Bunch--Davies vacuum. A generalization to other choices of vacuum was proposed in \cite{Cespedes:2020xqq}. The underlying principles is that time evolution in the bulk is implemented by a unitary transformation. 
Second, we review consequences of unitarity that leverage representation theory and are therefore specific to de Sitter spacetime and non-perturbative in nature. The simplest example is the de Sitter analog of the K\"all\'en--Lehmann representation of the two-point function (early discussions appeared in \cite{Bros:1994dn,Bros:1995js,Bros:1998ik}, and more recently in \cite{Hogervorst:2021uvp,DiPietro:2021sjt}). A more powerful result is the positivity of the spectral density appearing in the conformal partial wave decomposition of the four-point function, whose derivation borrows heavily from recent progress in the AdS/CFT literature. 

\subsection{Cosmological Optical Theorem and Cutting Rules}
\label{sec:COT}

Consider QFT in a generic FLRW spacetime, which we assume to be spatially flat for simplicity. Let's assume that an initial state is chosen at past infinity and that time evolution is generated by a unitary operator. In perturbation theory, the analytic structure of the initial state is preserved by time evolution. More concretely, consider a scalar field that at linear order obeys the equation of motion
\begin{equation}\label{eq:phiEOM}
\phi'' +\dfrac{2a'}{a}\phi'+\left[ c_s^2(\eta)k^2+a^2(\eta)m^2(\eta)\right] \phi=0\,,
\end{equation}
where we allowed for a generic time-dependent mass $m(\eta)$ and speed of sound $c_s(\eta)$ as long as they become approximately constant in the asymptotic past. The most relevant and best studied case is that in which one chooses the Bunch--Davies vacuum, which corresponds to selecting the positive-frequency solution of this differential equation in the asymptotic past, $\phi^+\sim e^{i c_s k \eta}$. This initial condition is ``Hermitian analytic" for complex $k$, namely
\begin{align}
    (e^{i c_s (-k^\ast) \eta})^\ast=e^{i c_s k \eta}\,.
\end{align}
Since the equation of motion is real, it can be proven that this property is maintained in the full solution \cite{Goodhew:2021oqg}
\begin{align}\label{disc1}
   \phi^+(k,\eta)=\phi^+(-k^\ast,\eta)^\ast \implies \disc[\phi] \equiv \phi(k)-\phi^\ast(-k^\ast)=0\,.
\end{align}
The calculation of wavefunction coefficients $\psi_n$ in~\eqref{eq:wfcoeff} in perturbation theory is organized in terms of Feynman diagrams that consist of a series of nested time integrals over bulk-to-bulk $G$ and bulk-to-boundary $K$ propagators. These are fixed by the mode functions above and hence inherit the property of Hermitian analyticity
\begin{equation}
\begin{aligned}
   K(-k^\ast,\eta)^\ast &=K(k,\eta)\,, \\
   G(-k^\ast,\eta)^\ast&=G(k,\eta)\,.
   \end{aligned}
\end{equation}
As proven in \cite{Melville:2021lst}, this leads to infinite set of propagator identities that relate the imaginary part of products of $K$'s and $G$'s to other products of $K$'s and fewer $G$'s. Assuming that all coupling constants in theory are real---as required by unitarity---and that all interactions are local, and hence Hermitian analytic, these propagator identities can be commuted with the time integrals over the local of interactions. This leads to infinitely many identities relating different combinations of wavefunction coefficients, which are called cosmological cutting rules. There is one such cosmological cutting rule per diagram to all orders in the perturbative expansion. 

\vskip4pt
In words, the cosmological cutting rules say that the sum over all possible ``cuts" of a diagram must vanish. It is often helpful to isolate the term in this sum with zero cuts, in which case one finds the schematic relation
\begin{align} \label{same}
   i \, \underset{\substack{ \text{internal} \\ \text{lines} } }{\disc }  \left[ i \, \psi^{(D)} \right]     &=  \sum_{\rm cuts}   \left[ \prod_{\substack{\rm cut \\ \rm momenta}} \int P \right]
   \prod_{\rm subdiagrams} (-i) \underset{\substack{ \text{internal } \& \\  \text{ cut lines} } }{\disc } \left[ i \, \psi^{(\rm subdiagram)} \right]\,,
\end{align}
where $D$ represents a diagram that is divided into subdiagrams by all possible cuts of one or more internal lines. An integral must be performed over every cut line including a factor of the associated power spectrum $P$. The $\disc$ acts as in~\eqref{disc1} and analytically continues the energies of all lines except those indicated in its subscript argument and with a minus sign on all spatial momenta
\begin{equation}
\begin{aligned}
 &  \underset{ k_1 \cdots k_j  }{\disc } f(k_1,\cdots,k_n;\{p\};\{\bfk \})\\
 &\hspace{1.25cm}
 \label{defdisc}\equiv f(k_1,\cdots,k_n;\{p\};\{\bfk \}) - f^\ast(k_1,\cdots,k_j,-k_{j+1},\cdots,-k_n;\{p\};-\{\bfk \}) \,, 
\end{aligned}
\end{equation}
\begin{figure}[t!]
    \centering
    \includegraphics[width=.95\textwidth]{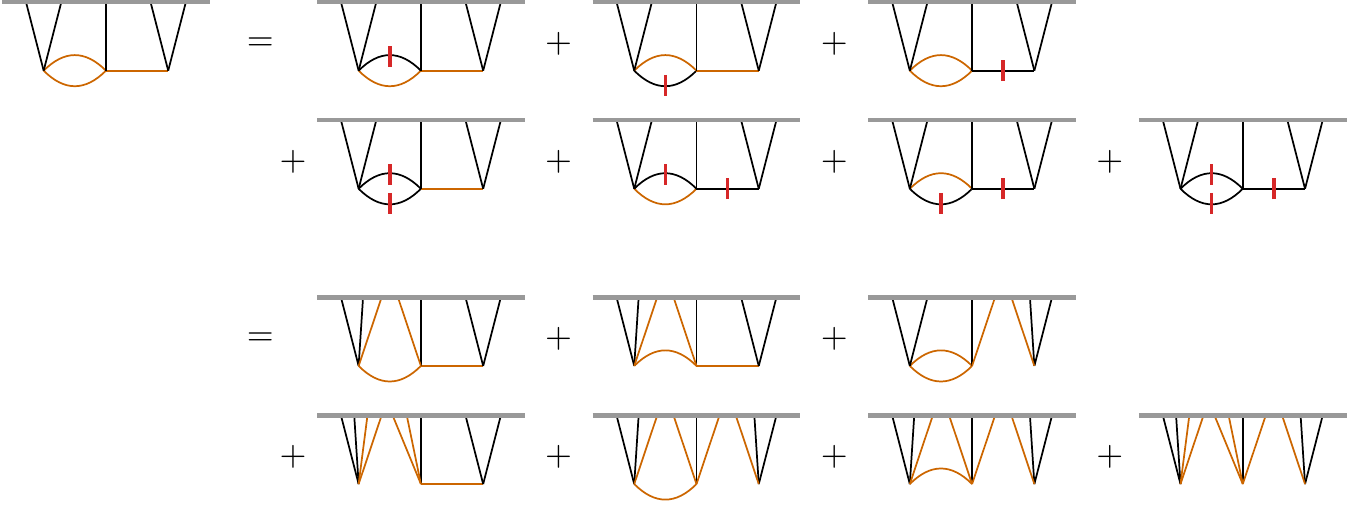}
    \caption{Graphical summary of the cosmological cutting rules for wavefunction coefficients~\cite{Melville:2021lst}. The discontinuity of a given diagram in perturbation theory with all internal energy kept fixed is equal to the product of the discontinuities of all disconnected diagrams obtained by summing over all possible ways to cut one or more internal legs and substituting them with a pair of external legs. For each cut leg, one should multiply by the associated power spectrum and integrate over the cut momentum. }
    \label{fig:cutting}
\end{figure}
\noindent
In Fig.~\ref{fig:cutting} we provide a graphical summary of these cutting rules. They can be summarized as follows. Consider a given diagram representing a specific contribution to an $n$-point wavefunction coefficient at some order in the perturbative expansion in coupling constants. We parameterize the kinematics by the $n$ external energies $\{k_1,k_2,\cdots,k_n\}$, all the energies of internal lines, and finally all rotation invariant products of momenta, such as $\v{k}_a\cdot \v{k}_b$. Then, sum over all possible ways to cut any number of internal lines. A ``cut" means substituting an internal line (bulk-to-bulk propagator) connecting two vertices with a pair of external lines (bulk-to-boundary propagators) attached to each of the vertices and a factor of the power spectrum of the cut momentum. This in general breaks up a diagram into a number of disconnected components. Take the discontinuity of each of these components separately keeping constant the energies of all internal lines and all cut lines. Finally, integrate over the momenta of all cut lines. The cutting rules dictate that the sum over all these terms must vanish. Notice that one can isolate the term with zero cut as in \eqref{same} and interpret the cutting rules as fixing the discontinuity of a given diagram in terms of that of other diagrams with fewer loops and/or fewer external legs. Relations of this type---between different orders in perturbation theory---are typical of unitarity. 

\vskip4pt
The above derivation is reminiscent of that of Cutkosky's cutting rules for amplitudes~\cite{Cutkosky:1960sp}, and in fact those relations should emerge on on the residue of the total energy pole, $E \to 0$, where wavefunction coefficients are related to Espace amplitudes~\cite{Meltzer:2020qbr}. In that case, one can think of cutting rules and encoding the content of the optical theorem to each order in perturbation theory. Analogously, we can refer to the collective constraints coming from the cosmological cutting rules as a {\it cosmological optical theorem} (COT). An open problem is that of finding a non-perturbative formulation of such a result in terms of the full wavefunction.

\vskip4pt
The COT is a powerful bootstrapping tool. To elucidate the cosmological cutting rules summarized above and to show how they can be used in practice to bootstrap new results we discuss below two examples: partial-energy recursion relations for tree-level exchange diagrams and the reconstruction of loop diagrams from their discontinuity.


\subsection{Partial-Energy Recursion Relations}\label{PERR}

The cosmological optical theorem is a powerful tool to bootstrap tree-level exchange diagrams without any assumption about dS boosts \cite{Jazayeri:2021fvk}. The general idea is to use the knowledge of the analytical structure of wavefunction coefficients to compute them via Cauchy's integral formula. Recall from Section~\ref{sec:correlators} that tree-level wavefunction coefficients have a very simple analytic structure: the only singularities are poles where partial energies vanish. In Minkowski, these are all simple poles so the amplitude limit in \eqref{kTeq0} is sufficient to fix all relevant residues \cite{Arkani-Hamed:2017fdk}. Conversely, in general FLRW spacetimes and in dS poles can have any order as in \eqref{psumA}. The crucial insight is that the cosmological optical theorem fixes the residues of all partial energy singularities of a given diagram in terms of lower-order diagrams \cite{Jazayeri:2021fvk}, in a way analogous to factorization theorems for amplitudes. What is left is fixing the terms that are regular in all partial energies, which appear as boundary terms in Cauchy's integral formula. This can be achieved by imposing the MLT (see Section~\ref{sec:MLT}).

\vskip 4pt
A good example of this procedure is the tree-level four-point function of four identical scalars $\psi_4$ from a single exchange, which we take here to be in the $s$-channel with the other channels simply obtained by permutations. Interactions are allowed to contain any number of time derivatives and local, rotation invariant contraction of spatial derivatives. The kinematic variables are the four external energies $\{ k_1,k_2,k_3,k_4\}$ plus the exchanged energy $k_I\equiv |\v{k}_1+\v{k}_2|$ and are conveniently arranged into the two partial energies $E_{L,R}$ and the total energy $E$:
\begin{align}
        E_L=k_1+k_2+k_I\,,\qquad E_R=k_3+k_4+k_I\,, \qquad E=k_1+k_2+k_3+k_4\,.
\end{align}
Inspired by recursion relations for amplitudes, the idea is now to extend $\psi_4$ to a function of a single complex variable $\tilde \psi_4(z)$ such that (\textit{i}) $\tilde \psi_4(0)=\psi_4$, (\textit{ii}) $\tilde \psi_4(z)$ is analytic in $z$ except for poles and (\textit{iii})  the residues of all the poles are fixed by unitarity. Such a function can be constructed with the following partial energy shift \cite{Benincasa:2018ssx,Jazayeri:2021fvk}\footnote{We omit the kinematical variables $k_1 k_2$ and $k_3 k_4$, as they don't change when applying the optical theorem.} 
\begin{align}
    \psi_4(E_L,E_R,k_I)\to \tilde{\psi}_4(z)=\psi_4(E_L+z,E_R-z,k_I)\,.
\end{align}
which is carefully crafted to avoid total energy poles, $k_T=0$, whose residues are not fixed by unitarity. The cosmological optical theorem for the tree-level $\psi_4$ dictates 
\begin{equation}
\begin{aligned}\label{ress}
      &\psi_4(E_L,E_R,k_I)+\psi^{\ast}_4(-E_L+2k_I,-E_R+2k_I,k_I)\\ 
      &\hspace{0.75cm}= P(k_I) \Big(\psi_3(E_L,k_I)-\psi_3(E_L-2k_I,-k_I)\Big) \Big(\psi_3(E_R,k_I)-\psi_3(E_R-2k_I,-k_I)\Big)\,.
\end{aligned}
\end{equation}
Notice that on the left-hand side only $\psi_4$ is singular at $E_{L,R} \to 0 $ and so in that limit the residues of all partial-energy poles are fixed by the right-hand side. We can now use Cauchy's integral formula to compute $\psi_4$ as a complex integral of $\tilde \psi_4(z)/z$ which is given by a sum over residues fixed by \eqref{ress} plus a boundary term that can be fixed by locality in the form of the MLT \cite{Jazayeri:2021fvk}. For some given cubic interactions, this procedure was shown to give a $\psi_4 $ that differs from that obtained from an {\it in-in} calculation only by contact interactions, as expected from the fact that these have vanishing discontinuity. More generally, this can be used to bootstrap all tree-level diagrams from lower-orders ones using only unitarity and locality, without any assumption about de Sitter boosts.


\subsection{Nonperturbative Implications of Unitarity}
\label{subsec::NPunitarity}

In the discussion of Section~\ref{sec:COT}, unitarity served to constrain the analytic structure of perturbative observables, essentially requiring that the structure present in the initial conditions was conserved as time evolves.  We now wish to discuss a different---but related---manifestation of unitarity: the positivity of the Hilbert space norm. 
Assuming invariance under the full dS isometry group, it is possible to derive a cosmological K\"all\'en--Lehmann (KL) representation of the two-point function and a conformal partial wave decomposition of the four-point function. In both cases unitarity implies the positivity of the corresponding spectral densities.


\vskip6pt
\noindent
{\bf Cosmological KL representation}\\[2pt]
\noindent
To begin with, we consider the two-point function~$\ex{\phi(\v{x},\eta)\phi(\v{x}',\eta)}$ in de Sitter space, where we take $\phi$ to be a scalar for simplicity.
This can only depend on the de Sitter invariant distance, $\xi$, between the points $\{\v{x},\eta\}$ and $\{\v{x}',\eta'\}$, which in planar coordinates~\eqref{equ:dS-Metric} takes the form 
\begin{equation}
    \xi\equiv\frac{4\eta \eta'}{|\v{x}-\v{x}'|^2-(\eta-\eta')^2}\,.
\end{equation}
We can insert a resolution of the identity between the two fields by summing over all possible unitary irreducible representations of the dS group. The two-point function then becomes a sum of terms with coefficients related to the norms of the intermediate states in each representation, which are required to be positive by unitarity. Putting this together, one finds~\cite{Bros:1995js}
\begin{equation}\label{KLrep}
    \ex{\phi(\v{x},\eta)\phi(\v{x}',\eta)}=\int_{\frac{d}{2}-i \infty}^{\frac{d}{2}+i\infty}{\rm d}\Delta\, \rho(\Delta)\, G(\xi,\Delta)\,,
\end{equation}
where $\rho(\Delta)\geq 0$ is the spectral density, while $G(\xi,\Delta)$ is the two-point function for a free scalar of mass $m$ with the Bunch--Davies vacuum choice. This can be written in terms of a hypergeometric function $_2F_1$, 
where $\Delta = \frac{d}{2}+i\mu$: 
\begin{equation}
    G(\xi;\Delta) = \frac{\Gamma(\frac{d}{2}+i\mu)\Gamma(\frac{d}{2}-i\mu)}{H^{1-d}(4\pi)^{\frac{d+1}{2}}\Gamma(\frac{d+1}{2})} \;_2F_1\!\left(\frac{d}{2}+i\mu,\frac{d}{2}-i\mu;\, \frac{d+1}{2};\, 1-\frac{1}{\xi}\right)\,.
\end{equation}
In~\eqref{KLrep}, we have only included the contribution from states in the principal series, but more generally the complementary and discrete series can contribute as well. This result is conceptually similar to the KL representation of the two-point function in Minkowski space: a general, two-point function can be written as an integral over free two-point functions with a positive spectral density. Upon taking the late-time limit this expression can be used to derive an inversion formula that computes the coefficient in a boundary operator expansion~\cite{Hogervorst:2021uvp}.

\vskip6pt
\noindent
{\bf Conformal partial wave expansion}\\[2pt]
\noindent
In perturbation theory, boundary correlators in de Sitter space, evaluated in the Bunch--Davies vacuum, share the same singularity structure as in Euclidean anti-de Sitter space (EAdS)~\cite{Sleight:2020obc,DiPietro:2021sjt,Sleight:2021plv}. This can be used to import insights from the EAdS setting into de Sitter space. (See Section~\ref{sec:AdS} for a discussion.) Boundary correlators in EAdS are single-valued functions of conformally invariant cross-ratios which, in turn, implies that they admit an expansion in terms of a special set of functions, ${\cal F}_{\Delta,J}$, known as \emph{conformal partial waves}.  
These form an orthogonal basis of single-valued eigenfunctions of the Casimir equation of the euclidean conformal group $SO\left(d+1,1\right)$.

\vskip4pt
Harmonic analysis on $SO\left(d+1,1\right)$~\cite{Dobrev:1977qv} implies that the partial-wave expansion of a single-valued conformally invariant four-point function of operators, ${\cal O}$, in $d$-dimensional euclidean space takes the form (say, in the (12)(34) channel): 
\begin{equation}\label{cpwe}
    \langle {\cal O}({\bf x}_1){\cal O}({\bf x}_2){\cal O}({\bf x}_3){\cal O}({\bf x}_4) \rangle={\mathds 1}_{12} {\mathds 1}_{34}+\sum\limits^\infty_{J=0}\int^{\frac{d}{2}+i\infty}_{\frac{d}{2}-i\infty} {\rm d}\Delta\,\rho_{J}(\Delta)\,{\cal F}^{12,34}_{\Delta,J}({\bf x}_1,{\bf x}_2,{\bf x}_3,{\bf x}_4)\,,
\end{equation}
where the first term is the contribution from the identity operator.
For boundary correlators in EAdS the spectral density $\rho_{J}(\Delta)$ is meromorphic as a function of $\Delta$, which is a consequence of the fact that the operator product expansion converges. 
In perturbation theory, dS boundary correlators can be written as a linear combination of EAdS Witten diagrams, which implies that $\rho_{J}(\Delta)$ is also meromorphic in dS space. 

\vskip 4pt
If the four-point function of interest continues to be single-valued at the nonperturbative level, then the expansion~\eqref{cpwe} will continue to hold nonperturbatively. This has been explored recently in~\cite{DiPietro:2021sjt,Hogervorst:2021uvp}. 
In terms of the conformal partial wave expansion~\eqref{cpwe}, unitarity in the  $SO\left(d+1,1\right)$ sense implies positivity of the spectral density for dS boundary correlators~\cite{DiPietro:2021sjt,Hogervorst:2021uvp}:
\begin{equation}\label{npunitaritydS}
    \rho_{J}(\Delta) \geq 0\,.
\end{equation}
The consequences of the positivity of this spectral density have not been fully explored, and such studies are interesting for the future. A particularly interesting open question is whether an analogue of the numerical conformal bootstrap can be formulated to constrain theories in de Sitter space. (See~\cite{Hogervorst:2021uvp} for some preliminary work in this direction.)

\section{From Anti-de Sitter to de Sitter}
\label{sec:AdS}

The structural similarities between de Sitter and anti-de Sitter space suggest that insights from the AdS setting can be imported into dS. This notion is buoyed by the fact that the natural questions to ask in AdS are also essentially holographic. Indeed, it has long been known that perturbative calculations in the two spaces are closely related. In this section, we summarize recent progress in utilizing this connection to develop new cosmological insights, and comment on the challenges to developing a holographic description of cosmology at the same level of refinement as that in AdS.

\subsection{Holography and Quantum Gravity}

In the search of a complete description of quantum mechanical and gravitational phenomena, we are inevitably led to consider observables on boundaries at infinity. On the one hand, only with an infinitely large apparatus are we free from errors due to quantum mechanical fluctuations of the apparatus itself. On the other hand, to avoid gravitational collapse the apparatus must be placed at the boundary of space-time. This is the common mantra that in quantum gravity there are no local observables and is closely related to the holographic principle, which suggests the existence of a purely boundary---or \emph{holographic}---description of the physics in the interior. 

\vskip 4pt
The AdS/CFT correspondence \cite{Maldacena:1997re,Gubser:1998bc,Witten:1998qj} provides an important working example of these ideas. It conjectures that quantum gravity in asymptotic anti-de Sitter space can be regarded as equivalent to a non-gravitational conformal field theory living on the boundary, which is flat Minkowski space. Lorentzian CFTs are, in particular, examples of theories where the rules are well understood. This is exemplified by the tremendous success of the conformal bootstrap program \cite{Hartman:2022zik}, which aims to carve out the space of CFTs simply by requiring basic quantum mechanical consistency. Remarkably, the requirements of conformal symmetry, unitarity and a consistent operator product expansion (crossing symmetry) has led to nontrivial bounds in the space of CFTs and the determination of critical exponents in the Ising and $O(N)$ models in three dimensions to record breaking accuracy~\cite{Kos:2016ysd}. In AdS, experiments that start and end at infinity are then computed by correlation functions of operators in the dual CFT description, meaning that the enigma of observables in quantum gravity in asymptotic anti-de Sitter space can be translated to sharp questions about the consistency of correlation functions in Lorentzian CFTs. This in turn has led to a wide range of powerful techniques to compute boundary correlators in AdS space which place consistency of the dual CFT description at the centre.

\vskip 4pt
It would be desirable to have a similar level of understanding for the universe we actually live in. To this end, a number of conceptual challenges need to be overcome. It should be noted that the success story of holography in AdS space largely stems from its causal structure. See Fig. \ref{fig:Penrose}. In AdS, the boundary lies at spatial infinity, meaning that the boundary theory is an ordinary quantum mechanical system with a standard notion of locality and time. Unitarity and causality of the bulk quantum gravity theory are therefore intimately related to unitarity and causality of the boundary quantum mechanics. This is to be contrasted with the situation in de Sitter space, where the role of time and space get interchanged: In dS, the boundaries are instead purely spatial and located at past/future infinity, which obscures how boundary correlators encode unitary time evolution in the interior of de Sitter space. In this type of scenario, dubbed ``time without time'', quantum mechanics itself should be an emergent concept hidden in some way in the structure of boundary observables.

\vskip 4pt
Under this map, inflationary backgrounds correspond in AdS holography to a class of slow-RG flows~\cite{Kaplan:2014dia}.  The symmetries characteristic of the boostless bootstrap for inflationary correlators correspond in (quasi)-AdS to a unitary, approximately scale-but-not-conformal field theory dual~\cite{Baumann:2019ghk}.  These analogues of inflationary background are consistent with known results about scale and conformal invariance in QFT~\cite{Nakayama:2013is}, but shows that natural questions about the nature of inflation can also spark new questions about AdS holography.

\vskip 4pt
The larger challenge for de Sitter holography (and the nonperturbative bootstrap more generally) is the lack of rigorously defined nonperturbative observables in de Sitter space in the presence of dynamical gravity~\cite{Bousso:1999cb,Witten:2001kn}.  As the metric fluctuates at the future boundary of de Sitter, one cannot define local boundary observables.  Calculations of the wavefunction can circumvent this problem because the boundary metric is fixed.  This motivates the dS/CFT approach to holography~\cite{Strominger:2001pn,Bousso:2001mw,Strominger:2001gp,Klemm:2001ea,Balasubramanian:2002zh,Maldacena:2002vr,Harlow:2011ke,Anninos:2011ui,Anninos:2012ft,Anninos:2012qw,Anninos:2017eib} which shares some features with the bootstrap approach to the wavefunction.  Unfortunately, the wavefunction approach only delays the problem, as the cosmological correlators of interest arise from integrating over the boundary metric. The full challenge of de Sitter holography includes dynamical gravity and remains an open problem in the cosmological slicing of de Sitter.  Approaches like dS/dS holography~\cite{Alishahiha:2004md,Alishahiha:2005dj,Dong:2018cuv,Gorbenko:2018oov} reflect this fundamental challenge, but are less directly connected to traditional cosmological observables.

\subsection{From AdS to dS and Back}

\begin{figure}[t!]
    \centering
    \includegraphics[scale=1]{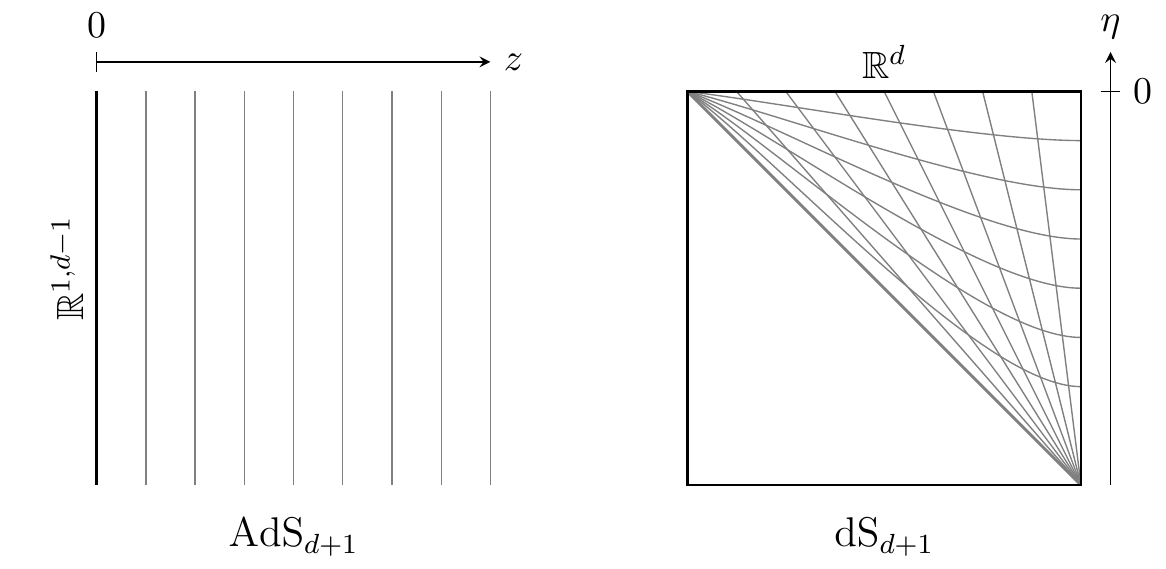}  \caption{Penrose diagrams showing the asymptotic boundaries of AdS ({\it left}) and dS ({\it right}). An important difference between the two cases is that the boundary in AdS is timelike, so that there is a boundary notion of causality and unitarity. On the other hand, the boundary of dS is spacelike, so the only natural notions of this kind are the ones inherited from the bulk spacetime, casting into relief the challenge of holography in this space.}
    \label{fig:Penrose}
\end{figure}

To bridge the gap between boundary correlators in AdS and dS, a natural starting point would be to try to understand the extent to which we can import our intuition from the AdS case. In both AdS and dS, the isometry group acts as the conformal group on the boundary: In AdS$_{d+1}$, this group is $SO(d,2)$ acting on the boundary $\mathbb{R}^{1,d-1}$, while, in dS$_{d+1}$, it is $SO(d+1,1)$ acting on $\mathbb{R}^{d}$ (see Fig.~\ref{fig:Penrose}). These can be placed on a similar footing by Wick rotating AdS$_{d+1}$ to $(d+1)$-dimensional Euclidean anti-de Sitter space (EAdS$_{d+1}$), which also has isometry group $SO(d+1,1)$ acting on the Euclidean boundary $\mathbb{R}^{d}$. Boundary correlators in (A)dS thus satisfy the same conformal Ward identities (reviewed in Section~\ref{sec:Symmetries}) and any differences in the way they encode consistent physics lies in the freedom left over after they are imposed. 

\vskip 4pt
As we have seen in Section~\ref{subsec::BIC}, the freedom remaining after imposing the conformal Ward identities is in the singularity structure: Unphysical singularities must be absent, while physical singularities must be normalised correctly. In the Bunch--Davies vacuum, what is regarded as an unphysical singularity is actually the same both in AdS and dS. In both cases folded singularities must be absent. In this case the difference between perturbative boundary correlators in EAdS and dS therefore solely lies in the physical singularities and their normalisation, which must be consistent with factorisation and unitarity in the respective space-times.

\vskip 4pt
The above suggests that, at least perturbatively, dS boundary correlators in the Bunch--Davies vacuum can be recast as boundary correlators in EAdS. Since unitarity manifests iself differently in dS and AdS, a priori the theory generating dS boundary correlators in EAdS is not necessarily the analytic continuation of a unitary theory in AdS. In the following we shall show this explicitly by using the fact that dS and EAdS are related by a double Wick rotation, to reduce the computation of dS boundary correlators in the {\it in-in} formalism to the computation of Witten diagrams in EAdS. In the next section this is revisited from a bootstrap perspective.

\vskip 4pt
 It will be useful to work in Poincar\'e coordinates, where the metric of EAdS reads 
\begin{equation}
 \text{d}s^2 = R^2_{\text{AdS}}\frac{\text{d}z^2+\text{d}{\bf x}^2}{z^2}\,.
\end{equation}
This is related to the flat slicing of the de Sitter metric in \eqref{equ:dS-Metric} via the following double Wick rotation:
\begin{equation}
    z = \pm i(-\eta)\,, \quad R_{\text{AdS}}=\pm iR_{\text{dS}}\,,
\end{equation}
where $R_{\text{dS}} \equiv 1/H$. At the level of correlation functions, the direction of the Wick rotation in $z$ should be specified to ensure the correct treatment of branch cuts. In particular, taking $z = + i(-\eta)$ has been shown to map AdS boundary correlators to wavefunction coefficients of the same theory in dS \cite{Maldacena:2002vr} and has been used to relate dS boundary correlators in the Bunch--Davies vacuum to their AdS counterparts \cite{McFadden:2009fg,McFadden:2010vh,McFadden:2011kk} at the level of three-point functions.

\vskip 4pt
One can also consider dS boundary correlators directly by relating the {\it in-in} formalism (see Section \ref{subsubsec::BttF}) to the computation of boundary correlators in EAdS \cite{Sleight:2020obc,Sleight:2021plv}. The $\pm$ branch of the {\it in-in} contour is obtained by the Wick rotation:
\begin{equation}
    \pm\,\text{branch}:\quad z = \left(-\eta\right)\,e^{\pm \frac{\pi i}{2}}, \label{ininWick}
\end{equation}
where the different branches are obtained by Wick rotating in opposite directions. Under \eqref{ininWick}, bulk-to-bulk propagators $G_{\Delta^\pm}$ in EAdS with Dirichlet $\left(\Delta^+\right)$ and Neumann $\left(\Delta^-\right)$ boundary conditions map to dS in-in propagators $G_{\pm\,{\hat \pm}}$ for the $\Delta^+$ and $\Delta^-$ modes.\footnote{Note the $\pm$ on $\Delta^\pm$ are unrelated to the $\pm$ branch of the {\it in-in} contour.} Solutions in the Bunch--Davies vacuum propagate the linear combination of $\Delta^+$ and $\Delta^-$ modes symmetric under $\Delta^+ \leftrightarrow \Delta^-$. Combined with \eqref{ininWick}, it follows that they can be expressed as the following linear combination of propagators $G_{\Delta^\pm}$ in EAdS:\footnote{For a detailed derivation of this identity, see Appendix~A.2 of \cite{Sleight:2021plv}.}  
\begin{align}\label{bubudSAdS}
    G^{\text{dS}}_{\pm\,{\hat \pm}}(\eta;{\bar \eta}) = c_{\Delta_+}\,e^{\mp \frac{i\pi \Delta_+}{2}}\,e^{{\hat \mp} \frac{i\pi \Delta_+}{2}}\,G^{\text{AdS}}_{\Delta^+}(z;{\bar z})+c_{\Delta_-}\,e^{\mp \frac{i\pi \Delta_-}{2}}\,e^{{\hat \mp} \frac{i\pi \Delta_-}{2}}\,G^{\text{AdS}}_{\Delta^-}\left(z;{\bar z}\right),
\end{align}
where $\pm$ and ${\hat \pm}$ refer, respectively, to the {\it in-in} contour branch of $\eta$ and ${\bar \eta}$. From \eqref{bubodSAdS} we see that taking a particle with scaling dimension $\Delta$ in EAdS to the $\pm$ branch of the {\it in-in} contour in dS entails multiplying by the phase $e^{\mp \frac{i\pi \Delta}{2}}$ while the coefficient $c_{\Delta}$ accounts for the difference in normalisation of the boundary two-point functions in (A)dS and can be found explicitly in \cite{Sleight:2021plv}, equation (2.15). Similarly, bulk-to-boundary propagators $K^{\text{dS}}_{\pm}$ on the $\pm$ branch are related to their EAdS counterpart $K^{\text{AdS}}_{\Delta}$ via 
\begin{equation}\label{bubodSAdS}
    K^{\text{dS}}_{\pm}\left(\eta\right) = c_{\Delta}\, e^{\mp \frac{i\pi \Delta}{2}}\,K^{\text{AdS}}_{\Delta}\left(z\right).
\end{equation}
The relations \eqref{bubudSAdS} and \eqref{bubodSAdS} between dS {\it in-in} propagators and their EAdS counterparts make clear that, perturbatively, any boundary correlator in the Bunch--Davies vacuum of dS can be expressed in terms of corresponding Witten diagrams in EAdS. 

\vskip 4pt
This mapping between AdS and dS correlators is simplest to illustrate at the level of contact diagrams, where the relation \eqref{bubodSAdS} between bulk-to-boundary propagators in (EA)dS implies that contact diagrams generated by the same vertex in dS and EAdS are proportional to each other. For example, for general $4$-point contact diagram, we have 
\begin{equation}
 \vcenter{\hbox{\includegraphics[scale=0.9]{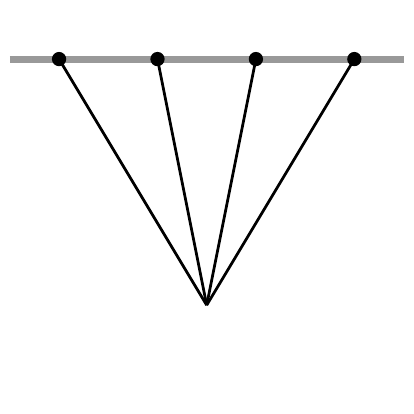}}} \quad    = \quad \lambda_{\Delta_1 \Delta_2 \Delta_3 \Delta_4} \ \  \vcenter{\hbox{\includegraphics[scale=0.9]{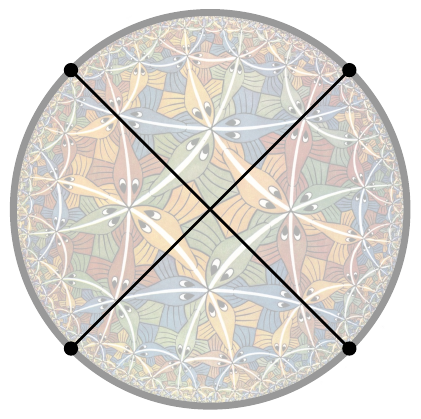}}}  
\end{equation}
The proportionality constant $\lambda_{\Delta_1\,\Delta_2\,\Delta_3,\Delta_4}$ itself can be determined using the analytic continuation \eqref{ininWick} and the relation \eqref{bubodSAdS}. For the $n$-point contact diagram generated by the non-derivative interaction $\phi_1 \phi_2 \ldots \phi_n$ of scalar fields $\phi_i$, by summing the contributions from the $\pm$ branches of the {\it in-in} contour one obtains \cite{Sleight:2019mgd}:
\begin{equation}\label{dsadscontact}
   \lambda_{\Delta_1\,\Delta_2\,\ldots\,\Delta_n} =   2\left(\prod\limits^n_{j=1} c_{\Delta_j} \right) \sin\left[ \left(\frac{d(n-2)}{4}+\tfrac{1}{2}\sum\limits^n_{j=1}\left(\Delta_j-\frac{d}{2}\right)\right)\pi\right] .
\end{equation}

\begin{figure}[htb]
    \centering
    \includegraphics[width=0.95\textwidth]{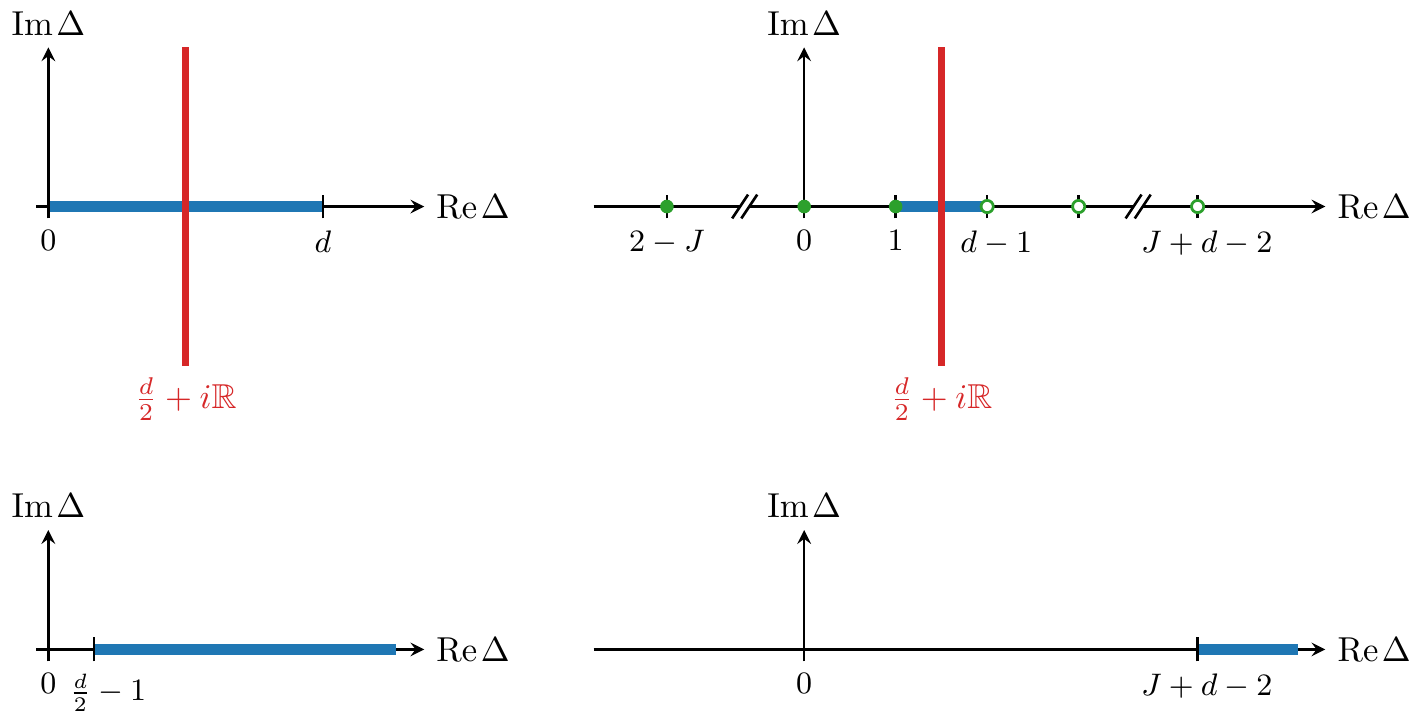}
    \caption{Plot of the unitary representations in de Sitter and anti-de Sitter space. {\it Top}: Unitary irreducible representations of the de Sitter group SO$(d+1,1)$ for scalars (left) and for spin-$J$ fields (right). {\it Bottom}: Unitary irreducible representations of the anti-de Sitter group SO$(d,2)$ for scalars (left) and spin-$J$ fields (right). Notice that the allowed representations are quite different in the two spaces, and are not simple analytic continuations of one another. For a more detailed discussion of (A)dS representations---and the differences between them---see e.g.~\cite{Dobrev:1977qv,Basile:2016aen,Sun:2021thf}.}
    \label{fig::representations}
\end{figure}

Note that the above relations between theories in dS and EAdS hold for arbitrary scaling dimension $\Delta$, i.e.~arbitrary masses of the corresponding bulk fields. However, as is well known, unitarity places restrictions on the values that the scaling dimensions can take. For theories in AdS, although we have Wick rotated to EAdS, we consider particles that are unitary irreducible representations (UIRs) of $SO\left(d,2\right)$, while in dS we consider UIRs of $SO\left(d+1,1\right)$. Crucially, these do not coincide!  See Fig.~\ref{fig::representations}. This implies that using the above relations to import from EAdS to dS may require input from a non-unitary theory in AdS.

\vskip 4pt
From the relations \eqref{bubudSAdS} and \eqref{bubodSAdS} between dS and EAdS propagators it  follows that it is possible to write down a Lagrangian in EAdS whose perturbative expansion matches that of the theory in dS \cite{DiPietro:2021sjt}. Consider a theory of a scalar field $\phi$ in dS,
\begin{equation}
    {\cal L}^{\text{dS}} = -\frac{1}{2}\partial_{\mu}\phi \partial^{\mu}\phi-\frac{1}{2}m^2 \phi^2-V^{\text{dS}}\left(\phi\right).
\end{equation}
The fact that dS propagators can be replaced by a linear combination of propagators for fields subject to the $\Delta^\pm$ boundary conditions in AdS tells us that the (perturbative) boundary correlators of this theory are equivalently reproduced by the following theory of two scalar fields $\Phi_{\Delta^\pm}$ in EAdS subject to the $\Delta^\pm$ boundary conditions: 
\begin{align}\label{nonUAdS}
    {\cal L}^{\text{AdS}}\left(\Phi_{\Delta^+},\Phi_{\Delta^-}\right) =\ & \sin\left(\pi(\Delta^+-\tfrac{d}{2})\right)\left(\partial_{\mu}\Phi_{\Delta^+}\partial^{\mu}\Phi_{\Delta^+}-m^2 \Phi_{\Delta^+}\Phi_{\Delta^+}\right)+\left(\Delta^+ \to \Delta^-\right) \nonumber \\ \nonumber &-e^{-i\pi (\frac{d-1}{2})}\,V^{\text{dS}}\big(e^{i\frac{\pi}{2} \Delta^+}\Phi_{\Delta^+}+e^{i\frac{\pi}{2} \Delta^-}\Phi_{\Delta^-}\big)\\&-e^{+i\pi(\frac{d-1}{2})}\,V^{\text{dS}}\big(e^{-i\frac{\pi}{2} \Delta^+}\Phi_{\Delta^+}+e^{-i\frac{\pi}{2} \Delta^-}\Phi_{\Delta^-}\big)\,. 
\end{align}
Note that the kinetic terms of the theory in EAdS are of incorrect sign, for all values of $\Delta^\pm$, which is another manifestation of the non-unitarity of an EAdS theory whose perturbative expansion generates dS boundary correlators. At the level of boundary correlators, this difference manifests itself in the relative coefficient \eqref{dsadscontact} between dS and EAdS contact diagrams. The latter in particular implies that contact diagrams in dS vanish for certain collections of particles in certain dimensions, which also follows from the cosmological optical theorem (see Section \ref{sec:COT} and the corresponding result below equation \eqref{WFtoCorrelator}).

\subsection{Bootstrapping Perturbative Correlators} 

In the previous section, we showed how boundary correlators in dS can perturbatively be expressed as a linear combination of Witten diagrams in EAdS. Formally, the precise relative coefficients of the EAdS Witten diagrams follow either from the {\it in-in} formalism or the non-unitary Lagrangian \eqref{nonUAdS}. In practice, however, it can be quite cumbersome to evaluate them beyond the simplest of diagrams and, moreover, since they require us to take various auxiliary steps (which individually are unphysical) they obscure the properties of the final result.

\vskip 4pt
To align with the bootstrap philosophy, we seek an approach to fix the dS boundary correlators that places their consistency at the centre. From this perspective, the fact that perturbative boundary correlators in the Bunch--Davies vacuum in dS can be expressed as a linear combination of corresponding EAdS Witten diagrams follows as a consequence of:
\begin{enumerate}
    \item {\it Symmetries}: In both cases, the boundary correlators must obey conformal symmetry, and in particular must obey the conformal Ward identities reviewed in Section \ref{sec:Symmetries}.

    \item {\it Initial conditions}: Throughout we have considered boundary correlators in the Bunch--Davies vacuum, which requires the absence of folded singularities \cite{Chen:2006nt,Holman:2007na,LopezNacir:2011kk,Flauger:2013hra,Aravind:2013lra}. Folded singularities are also absent in AdS boundary correlators owing their single-valuedness in the Euclidean region. 
\end{enumerate}
 Like for the bootstrap of perturbative dS correlation functions presented in Section \ref{subsec::BIC}, the remaining freedom is fixed by on-shell factorisation and the requirement that particles correspond to UIRs of the appropriate isometry group. Factorisation implies that the relative coefficients of the EAdS Witten diagrams appearing in a given dS boundary correlator are given by the product of coefficients \eqref{dsadscontact} that relate each contact subdiagram in dS to their EAdS counterpart \cite{Sleight:2021plv}. For example, the tree-exchange diagram of a particle mass \eqref{mass} in dS can be written as a sum of the two tree-exchange diagrams in EAdS for the $\Delta^\pm$ modes \cite{Sleight:2020obc}: 
\begin{align}\label{dSasAdSexch}
    \vcenter{\hbox{\includegraphics[scale=0.9]{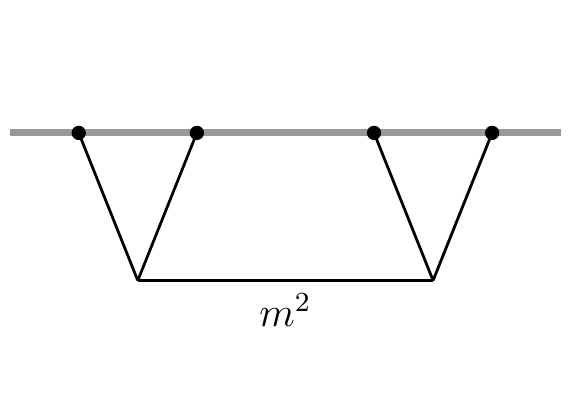}}}  &\quad = \quad \sum_{I = \pm}\  \frac{\lambda_{\Delta_I}^2}{c_{\Delta_I}} \ \  \vcenter{\hbox{\includegraphics[scale=0.9]{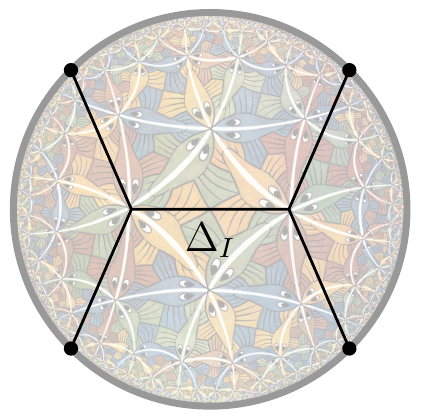}}} 
\end{align}
The coefficients $\lambda_{\Delta^\pm}$ convert the two three-point contact subdiagrams for each $\Delta^\pm$ mode in the AdS exchange to their dS counterpart as in \eqref{dsadscontact}. 

\vskip 4pt
More generally, starting from a given Witten diagram in EAdS with definite boundary conditions imposed on any exchanged particle, one can obtain the corresponding boundary correlator in dS in the Bunch--Davies vacuum as follows \cite{Sleight:2021plv}: ($i$) For each contact subdiagram, multiply by the factor \eqref{dsadscontact} which converts it to its dS counterpart; ($ii$) For each internal line with mode, say $\Delta$, divide by $c_{\Delta}$ accounting for the change in two-point function normalisation from AdS to dS. ($iii$) Symmetrise under the interchange of $\Delta^+$ and $\Delta^-$ boundary conditions for each internal line. 

\vskip 4pt
To summarise, in the case that the universe at early times was in the Bunch--Davies vacuum there is the potential to make significant progress in closing the gap between boundary correlators in AdS and dS. We have seen that, perturbatively, dS boundary correlators in the Bunch--Davies vacuum can be expressed as a linear combination of Witten diagrams generated by the same collection of particles and couplings in EAdS. The relative coefficients of the Witten diagrams encode perturbative unitarity in dS and ensure consistent on-shell factorisation. 

\vskip 4pt
Let us note that this result implies that in the Bunch--Davies vacuum boundary correlators in dS and EAdS have the same singularity/analytic structure. This opens up the possibility to leverage a variety of powerful techniques that were originally developed in the context of the conformal bootstrap approach to boundary correlators in EAdS (see e.g. \cite{Bissi:2022mrs} for a recent review) to study dS boundary correlators---at least those which do not rely on unitarity in AdS. An example was presented in Section \ref{subsec::NPunitarity} when considering the conformal partial wave expansion of dS boundary correlators, which assumes that they are single-valued. Since boundary correlators in AdS are single-valued in the Euclidean region, the above results establish that---at least perturbatively---the same is true in dS \cite{Sleight:2020obc,DiPietro:2021sjt,Sleight:2021plv}.


\section{Conclusions and Outlook}
\label{sec:Conclusions}

\begin{figure}[b!]
    \centering
    \includegraphics[scale=1]{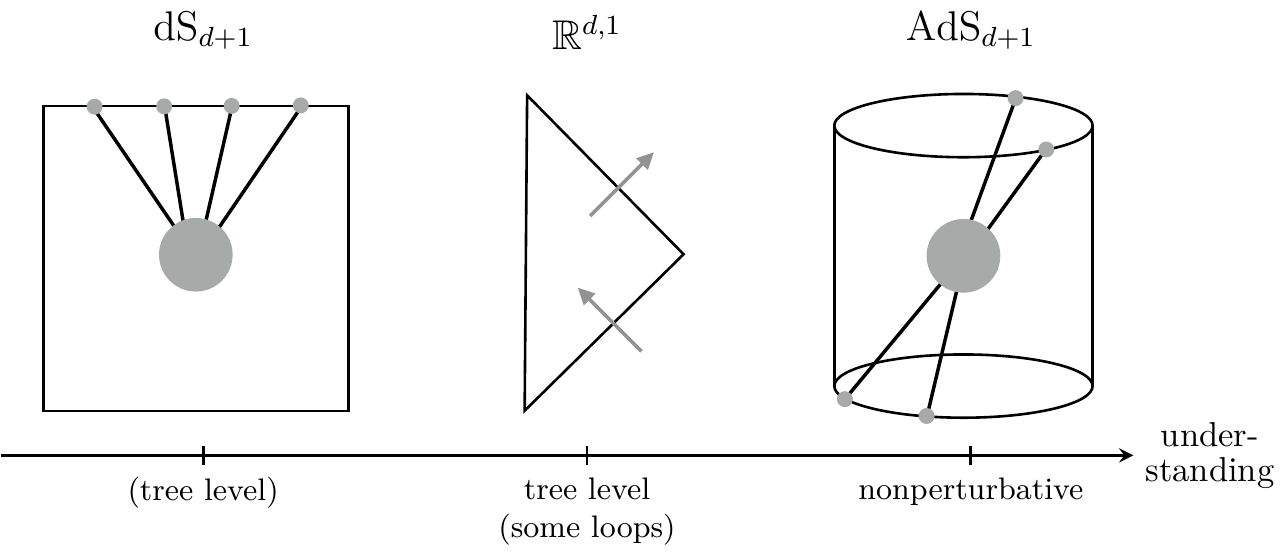}
    \caption{Remarkably, the closer we get to spacetimes that describe the universe we live in (de Sitter space), the less we understand about how to rigorously define a theory in that space. In contrast, we have a perfect nonperturbative definition of quantum gravity in AdS spacetimes in terms of a boundary CFT. In flat space, we have some understanding of the perturbative structure of QFT and partial understanding of some nonperturbative effects. On the other hand, we are just beginning to scratch the surface of the rules underlying consistent theories in dS, where our understanding is not even complete at tree level.}
    \label{fig:Conclusions}
\end{figure}

The bootstrap approach to cosmological correlations is still nascent. 
There is an irony to the fact that the spacetimes most similar to our own---asymptotically de Sitter---are still the most mysterious (see Fig.~\ref{fig:Conclusions}). 
In recent years, progress has been made in many directions, some of which we have reviewed. Nevertheless, there are still many puzzles to decipher and challenges to overcome. These challenges provide opportunities for further progress, and we close by listing some important open directions to pursue in the coming years.

\begin{itemize}
\item {\bf Beyond Feynman diagrams:}\hspace{0.3cm}The current guise of the cosmological bootstrap still somewhat mirrors bulk perturbation theory, proceeding diagram by diagram. However, the most dramatic manifestations of the simplifying power of the S-matrix bootstrap reveals themselves in situations where observables cannot be meaningfully split into various diagrammatic contributions, for example in gauge theories and gravity. Indeed, one could view the discovery of the Parke--Taylor formula~\cite{Parke:1986gb}---a dramatic simplification of tens of thousands of Feynman diagrams---as marking the beginning of the modern amplitudes revolution. Similar simplifications have not yet been achieved in the cosmological setting. However, given that scattering amplitudes live within cosmological correlators, it is natural to expect that these structures exist, and finding them is an important goal. Whenever cosmology's Parke--Taylor moment does arrive, we expect that it will catalyze a number of new discoveries.

\item
{\bf From trees to loops:}\hspace{0.3cm}Most of the explicit results in the cosmological bootstrap have been derived for tree-level correlators. It is important both conceptually and practically to push our understanding further in perturbation theory. As we reviewed in Section~\ref{sec:Symmetries}, we now have a reasonably complete understanding of the singularity structure at tree level. However, aside from isolated examples, at loop level little is known in a systematic fashion. Developing a similar understanding at one-loop to what is currently know at tree level presents a concrete challenge that can serve as a gateway to better understanding the analytic structure of correlation functions more generally in perturbation theory. Beyond this, there are a number of situations where loop corrections are practically important. For example, from the phenomenology side, the leading couplings of the inflaton to fermions (including Standard Model fields) only arise at one-loop~\cite{Green:2013rd,Chen:2016uwp,Chen:2016hrz,McAneny:2017bbv}. Moreover, there has been much recent work exploring both the stability of de Sitter space~\cite{Palti:2019pca} (and the consistency of perturbation theory in cosmological spacetimes~\cite{Ford:1984hs,Antoniadis:1985pj,Tsamis:1994ca, Tsamis:1996qm,Tsamis:1997za,Polyakov:2007mm,Polyakov:2009nq,Senatore:2009cf,Giddings:2010nc,Giddings:2010ui,Burgess:2010dd,Marolf:2010nz,Krotov:2010ma,Marolf:2010zp,Rajaraman:2010xd,Marolf:2011sh,Giddings:2011zd,Giddings:2011ze,Senatore:2012nq,Pimentel:2012tw,Senatore:2012ya,Polyakov:2012uc,Beneke:2012kn,Akhmedov:2013vka,Anninos:2014lwa, Akhmedov:2017ooy,Hu:2018nxy,Akhmedov:2019cfd,Gorbenko:2019rza,Baumgart:2019clc,Mirbabayi:2019qtx,Cohen:2020php,Mirbabayi:2020vyt,Baumgart:2020oby,Cohen:2021fzf}) and the dynamics of eternal inflation~\cite{Vilenkin:1983xq,Linde:1986fd,Linde:1986fc,Arkani-Hamed:2007ryv,Creminelli:2008es,Leblond:2008gg,Dubovsky:2008rf,Brahma:2019iyy,Rudelius:2019cfh,Cohen:2021jbo}, which are both situations where loops become important, and further developing the bootstrap to approach these cases will also provide opportunities to make connections with this work.

\item
{\bf Uncovering hidden structures:}\hspace{0.3cm}The deepest and most far-reaching insights arising from the S-matrix bootstrap have involved the discovery of completely unexpected physical and mathematical structures~\cite{Elvang:2013cua}. 
In many cases these structures were uncovered through the change in perspective provided by the on-shell philosophy.
Of course, it is difficult to plan to discover similar such hidden structures in cosmology. 
Fortunately, we can leverage the successes of the S-matrix and conformal bootstrap for some clues of where to look. 
As an example, one of the more surprising structures lurking inside scattering amplitudes are double copy relations. First noticed within string theory~\cite{Kawai:1985xq}, they are now known to be much more far-reaching in field theory, not only allowing gravity amplitudes to be expressed as suitably-understood squares of Yang--Mills amplitudes~\cite{Bern:2008qj}, but more generally connecting a web of theories~\cite{Cachazo:2014xea}. 
An interesting concrete challenge is to develop a similar understanding in the cosmological context. Some progress has been made recently~\cite{Li:2018wkt,Farrow:2018yni,Fazio:2019iit,Lipstein:2019mpu,Albayrak:2020fyp,Armstrong:2020woi,Diwakar:2021juk,Jain:2021qcl,Herderschee:2022ntr,Sivaramakrishnan:2021srm,Cheung:2022pdk}, but the fate of the cosmological double copy remains somewhat mysterious. 
We expect that resolving this mystery will reveal that the double copy is just the tip of the iceberg, and that there are other beautiful structures within cosmological correlators waiting to be mined.

\item
{\bf Pushing the limits of EFT:}\hspace{0.3cm} The majority of developments thus far have focused on situations where the inflationary background is close to de Sitter space (though allowing for sizeable breaking of de Sitter symmetries in interactions) and low-order correlation functions are the dominant signature~\cite{Snowmass2021:CosmoEFT}. However, large deviations from these implicit assumptions are not excluded by observations, and suggest novel observables like oscillatory features~\cite{Slosar:2019gvt,Snowmass2021:InfTH}. These situations are less symmetric and thus present a novel challenge for the bootstrap approach.  For similar reasons, the signatures themselves vary significantly from model to model and thus finding an organizing principle for these less symmetric cases is of broad interest. The tail of the distribution of density fluctuations~\cite{Panagopoulos:2019ail,Vennin:2020kng,Achucarro:2021pdh,Celoria:2021vjw} and/or higher N-point functions~\cite{Panagopoulos:2020sxp} are areas where novel theoretical insights could have important observational consequences.  Enhanced information at large multiplicity can arise in models of cosmological particle production~\cite{Flauger:2016idt}, but may be better understood as a feature in a map rather than a change to the statistics~\cite{Munchmeyer:2019wlh,Baumann:2021ykm}.  There has been much recent interest in flat space at large-multiplicity perturbation theory and non-trivial classical saddles (e.g.~\cite{Monin:2016jmo,Badel:2019oxl}). It will be very interesting to understand these features more systematically and their cosmological implications. Often we learn about a formalism by pushing it to its limits, and trying to approach these situations where the EFT naively breaks down will doubtless be illuminating. 

\item
{\bf From IR to UV:}\hspace{0.3cm}In the inflationary paradigm, the largest-scale structures we see today originated as quantum-mechanical fluctuations in the very early universe. This striking bridge between large and small is a remarkable feature of inflation, and a unique opportunity to learn about physics at very short distances.
To maximally leverage this connection, it is critical to make the link between infrared observables and ultraviolet physics completely precise.
In the context of scattering amplitudes, our understanding of the structure of the S-matrix is mature enough that we can connect the IR physics that we observe, and the UV physics that we want to know about, by means of powerful dispersion relations and positivity bounds~\cite{deRham:2022hpx}. 
Deriving analogous relations in cosmology requires two main ingredients: 1) an improved (nonperturbative) understanding of the consequences of unitarity for cosmological correlators and 2) further insights into their analytic structure. Fortunately, these developments are already underway. 
As a concrete first step, it will be important to connect our understanding of perturbative and nonperturbative unitarity in de Sitter space. Further development of perturbative techniques at loop level and beyond will provide insights necessary to characterize the analytic structure more fully. Another important direction to pursue is to elucidate the consequences of de Sitter causality for cosmological correlators, which has not been extensively studied.
These insights can then be synthesized into dispersion relations that will constrain cosmological EFTs, which will be an important milestone in the development of the cosmological bootstrap.

\item {\bf Carving out theory space:}\hspace{0.3cm}An important goal of the cosmological bootstrap is to classify the space of consistent cosmological field theories. There are two aspects to this broad theme. One is to understand in a fixed cosmological background what QFTs can consistently be defined. The second and more ambitious goal is to go one step further and classify the full space of models that can give rise to inflation in the first place. In the perturbative context, some consistency requirements are already known, and it is important to fully map out the space of consistent theories. 
It is also critical to further develop nonperturbative bootstrap tools that constrain the space of possible inflationary models.
An important question is whether ``single field" inflation can occur in a consistent model of quantum gravity. That is, can the inflaton be an isolated degree of freedom, with a parametric gap to other states? Or are there necessarily other fields that are important to the inflationary dynamics? A famous fact about de Sitter space is that it does not admit linearly realized supersymmetry~\cite{Pilch:1984aw,Lukierski:1984it,Anous:2014lia}. Hence, one might speculate that it will be difficult or impossible to find de Sitter solutions in string theory with a parametric gap to the string scale if indeed supersymmetry plays a fundamental role. Fortunately, this is a question that can be approached systematically utilizing tools developed in the context of the conformal bootstrap~\cite{Hartman:2022zik}, which have already begun to be imported into the cosmological setting~\cite{Sleight:2020obc,DiPietro:2021sjt,Hogervorst:2021uvp}. 

\item {\bf Towards the Veneziano correlator:} The inception of string theory traces its way back to the discovery of the Veneziano amplitude. From this remarkable structure, innumerable marvels have emerged. In de Sitter space, one can place sharp requirements on the properties of the analogous ``Veneziano correlator". Such a UV-complete correlator would not have the field-theoretic energy singularities described in Section~\ref{sec:sing} and would posses a Regge-like spectrum of resonances with corresponding oscillatory factors analogous to~\eqref{eq:massscalesqueezed}. If such an object exists, it would provide a glimpse into string theory in de Sitter space in the regime where there is no separation between the string scale and the Hubble scale. Such a scenario would be extremely interesting because these string states would be excited during inflation leading to striking signatures.

\item
{\bf De Sitter holography:}\hspace{0.3cm}One of the deepest goals of cosmology as a discipline is to uncover what really happened at the initial singularity in the early universe. This is a question for quantum gravity in cosmological spacetimes. Given our experiences in AdS---and with quantum gravity more generally---it stands to reason that this is a question best addressed holographically. However, holography in de Sitter space is underdeveloped compared to AdS. 
Conceptually, the difficulty compared to anti-de Sitter space is that the holographic direction in the de Sitter context is {\it time}, and so our familiar notions of unitarity and causality must be  emergent in this setting. A further challenge is that in the presence of dynamical gravity, the future boundary itself fluctuates, making it harder to sharply define observables. Notably, none of the developments described in this paper posit the actual existence of a holographic dual to inflation or cosmology (though there is much interesting work in this direction, e.g.,\cite{Strominger:2001pn,Bousso:2001mw,Strominger:2001gp,Klemm:2001ea,Balasubramanian:2002zh,Maldacena:2002vr,Harlow:2011ke,Anninos:2011ui,Anninos:2012ft,Anninos:2012qw,Anninos:2017eib}), instead they are relying on kinematic properties of boundary correlators, or on other consequences of bulk dynamics.
Nevertheless, these perturbative results serve as a sort of theoretical data, which any putative holographic dual must reproduce,
and also help us understand the rules that such a boundary theory must obey. 
There is evidence that it should share some properties with a Euclidean CFT, though it is unclear what the allowed spectrum of such a theory is or should be. Despite the obvious challenges, we view this as an essential open problem in cosmological physics, and it provides an opportunity for connection with other efforts in the Theory Frontier~\cite{Hartman:2022zik,Faulkner:2022mlp,Bousso:2022ntt,Snowmass2021:InfTH}.

\item
{\bf New observational strategies:}\hspace{0.3cm}A central objective of the cosmological bootstrap is to understand how fundamental physical principles are encoded in cosmological observables. This more refined understanding of the signatures of fundamental physics will suggest observational targets to look for in data, or protected observables that cannot be mimicked by late-universe effects. There are already several success stories of this philosophy~\cite{Baumann:2017lmt,Green:2020whw,Baumann:2021ykm} and further developing this line of inquiry is extremely important to be able to fully utilize the inflationary epoch as a tool for discovery, and to decode the physics of inflation itself.

\end{itemize}

\noindent
The question of the origin of the universe is one that has captured human imagination for centuries. 
We are lucky enough to live in a time where this is a scientific question that can be approached systematically.
Nevertheless, information about the universe's earliest moments is scant---we cannot directly image this epoch, instead we must reconstruct its history from subtle correlations frozen on its boundary.
This difficulty is an opportunity in disguise, as it challenges us to make sense of the time evolution of the universe from this static vista.
To guide ourselves, we fall back on cherished fundamental principles in order to understand how to reconstruct observables directly on this late-time surface. 
The hope is that this will further illuminate the path forward.
In this white paper, we have summarized some recent progress in this direction. These are only the first steps and it is clear there is a long journey ahead, 
but one where many new discoveries surely await us.

\vskip 10pt
\paragraph{Acknowledgements} DB, AJ, and GP thank Nima Arkani-Hamed, Wei-Ming Chen, Carlos Duaso Pueyo, and Hayden Lee for collaboration on related topics.
DB receives funding from a VIDI grant of the Netherlands Organisation for Scientific Research~(NWO) and is part of the Delta-ITP consortium.  DB is also supported by a Yushan Professorship at National Taiwan University funded by the Ministry of Science and Technology (Taiwan). DG is supported by the US~Department of Energy under grant no.~\mbox{DE-SC0009919}. EP has been partially supported by the research program VIDI with Project No. 680-47-535, which is financed by the Netherlands Organisation for Scientific Research
(NWO) and by STFC consolidated grant ST/T000694/1. GP is supported by a grant of the Netherlands Organisation for Scientific Research~(NWO/OCW) and through the Delta-ITP consortium of NWO/OCW. CS is supported by STFC through grant ST/T000708/1. MT is supported by the INFN through the research initiative STEFI.

\clearpage
\phantomsection
\addcontentsline{toc}{section}{References}
\bibliographystyle{utphys}
{\linespread{1.075}
\bibliography{refs}
}

\end{document}